\documentclass[twocolumn]{aastex7}
\usepackage[utf8]{inputenc}
\usepackage[T1]{fontenc}
\usepackage{amsmath}
\usepackage{graphicx}

\renewcommand{\vec}[1]{\boldsymbol{#1}}

\defcitealias{besla_role_2012}{B12}
\newcommand{\Bref}{\citetalias{besla_role_2012}}

\shorttitle{BFEs of the Clouds}
\shortauthors{Foote et al.}
\graphicspath{{./}{figures/}}

\begin{document}

\title{Mapping the Distorted Dark Matter Distribution of the LMC-SMC System Prior to Milky Way Infall with Basis Function Expansions}

\correspondingauthor{Hayden R. Foote}
\author[orcid=0000-0003-1183-701X, sname=Foote, gname=Hayden]{Hayden R. Foote}
\affil{Steward Observatory, University of Arizona, 933 North Cherry Avenue, Tucson, AZ 85721, USA}
\email[show]{haydenfoote@arizona.edu}

\author[orcid=0009-0009-0158-585X, sname=Rathore, gname=Himansh]{Himansh Rathore}
\affil{Steward Observatory, University of Arizona, 933 North Cherry Avenue, Tucson, AZ 85721, USA}
\email{himansh@arizona.edu}

\author[orcid=0000-0003-0715-2173, sname=Besla, gname=Gurtina]{Gurtina Besla}
\affil{Steward Observatory, University of Arizona, 933 North Cherry Avenue, Tucson, AZ 85721, USA}
\email{gbesla@arizona.edu}

\author[orcid=0000-0001-7107-1744, sname=Garavito-Camargo, gname=Nicol\'as]{Nicol\'as Garavito-Camargo}
\thanks{NASA Einstein Fellow}
\affil{Steward Observatory, University of Arizona, 933 North Cherry Avenue, Tucson, AZ 85721, USA}
\affil{Department of Astronomy, University of Maryland, 4296 Stadium Drive,
College Park, MD 20742-2421}
\affil{Center for Computational Astrophysics, Flatiron Institute, Simons Foundation, 162 Fifth Avenue, New York, NY 10010, USA}
\email{garavito@umd.edu}

\author[orcid=0000-0002-9820-1219, sname=Patel, gname=Ekta]{Ekta Patel}
\affil{Department of Astrophysics and Planetary Science, Villanova University, 800 E. Lancaster Avenue, Villanova, PA 19085, USA}
\email{ekta.patel@villanova.edu}

\author[orcid=0000-0003-1517-3935, sname=Petersen, gname=Michael]{Michael S. Petersen}
\affil{Institute for Astronomy, University of Edinburgh, Royal Observatory, Blackford Hill, Edinburgh EH9 3HJ, UK}
\email{michael.petersen@roe.ac.uk}

\author[orcid=0000-0003-2660-2889, sname=Weinberg, gname=Martin]{Martin D. Weinberg}
\affil{Department of Astronomy, University of Massachusetts, Amherst, MA 01003-9305, USA}
\email{weinberg@astro.umass.edu}

\author[orcid=0000-0003-4232-8584, sname=G\'omez, gname=Facundo]{Facundo A. G\'omez}
\affil{Departamento de Astronom\'ia, Universidad de La Serena, Av. Raúl Bitrán 1305, La Serena, Chile}
\email{fagomez@userena.cl}

\author[0000-0003-3922-7336]{Chervin F. P. Laporte}
\affiliation{LIRA, Observatoire de Paris, Universit\'e PSL, Sorbonne Universit\'e, Universit\'e Paris Cit\'e, CY Cergy Paris Universit\'e, CNRS, 92190 Meudon, France}
\affiliation{Institut de Ciencies del Cosmos (ICCUB), Universitat de Barcelona (IEEC-UB), Martí i Franquès 1, E-08028 Barcelona, Spain}
\affiliation{Kavli IPMU (WPI), UTIAS, The University of Tokyo, Kashiwa, Chiba 277-8583, Japan}
\email{Chervin.Laporte@obspm.fr}

\collaboration{all}{The EXP Collaboration}

\begin{abstract}

The SMC orbits within the LMC's dark matter (DM) halo in a $\sim$1:10 mass-ratio encounter. The LMC:Milky Way (MW) interaction is also $\sim$1:10, and is expected to perturb the MW's DM distribution. However, no framework exists to quantify the severity of these perturbations over multiple pericenters and longer periods of time, such as the LMC-SMC interaction history. We construct basis function expansions of a high-resolution \textit{N}-body simulation of the Clouds interacting in isolation and analyze their DM distributions at an epoch approximating the time of their infall to the MW. Our goal is to quantify how the Clouds distort each other's DM distributions \textit{without} the MW. The LMC halo's response to the SMC includes a $\sim 20$ kpc long dynamical friction wake and the displacement of the LMC's density center during each SMC pericenter, which produces two overdensities in the LMC halo (at $\sim$60 and $\sim$100 kpc) at MW infall. The SMC's tidal radius at infall is just $\sim4$ kpc, at which point the SMC has lost two-thirds of its initial DM mass to the LMC. The distortions to the Clouds' halos produce a highly asymmetric acceleration field. Accurate orbit integration in the LMC-SMC system must account for the time-dependent shapes of both halos. The SMC-induced perturbations in the LMC DM halo resemble the MW-LMC system, and persist over multiple SMC pericenters. We conclude that 1:10 satellite-host encounters induce characteristic deformations in both DM halos across host-mass scales, with implications for merger rates and tests of DM models.

\end{abstract}



\section{Introduction}\label{sec:intro} 

The LMC and SMC (collectively the ``Clouds'') are the Milky Way (MW)'s most massive interacting pair of satellites. Due to their proximity ($\sim$50 kpc), a wealth of existing spectroscopic, astrometric, and photometric data are available for the Clouds, making them an ideal laboratory for dark matter (DM) physics. These data have been used to demonstrate that mutual tidal interactions between the Clouds can influence their baryonic morphology (e.g., \citealt{besla_role_2012}, hereafter \Bref{}; \citealt{mackey_10_2016, cullinane_magellanic_2022}), kinematics \citep[e.g.,][]{choi_recent_2022, munoz_chemo-dynamical_2023, rathore_response_2025, vijayasree_vmc_2025}, star formation histories \citep[e.g.,][]{harris_star_2009, weisz_comparing_2013, massana_synchronized_2022}, and form the Stream of HI gas trailing the Clouds \citep[e.g. reviews by][]{donghia_magellanic_2016, lucchini_following_2024}. However, no framework currently exists to quantify the distortions in the Clouds' DM halos, which is ultimately necessary to unlock their potential to test DM physics. The goal of this paper is to develop such a framework and to generalize our findings to 1:10 mass ratio encounters more broadly.

Models of the MW-LMC system have already proven the power of this approach. As the MW's most massive satellite ($\sim 1:10$ mass ratio; e.g. \citealt{garavito-camargo_hunting_2019} and references therein), the LMC has thrown the MW into a state of disequilibrium, with far-reaching consequences for the MW's DM halo, stellar halo, and disk (see \citealt{vasiliev_effect_2023} for a recent review). 

In particular, idealized models predict that the LMC induces two major perturbations in the MW's DM halo: a local dynamical friction wake trailing the LMC and a density dipole owing to the displacement of the MW's density center as the LMC causes the MW's inner halo to move with respect to its outer halo \citep{gomez_and_2015, garavito-camargo_hunting_2019, petersen_reflex_2020, garavito-camargo_quantifying_2021, rozier_constraining_2022, lilleengen_effect_2023, vasiliev_dear_2024}. These predictions are echoed in MW-LMC analogs from cosmological simulations \citep[]{arora_shaping_2025, darragh-ford_shaping_2025} and borne out by numerous observations of the MW's stellar halo \citep[e.g.,][]{belokurov_pisces_2019, conroy_all-sky_2021,  erkal_detection_2021, petersen_detection_2021, fushimi_determination_2024, amarante_mapping_2024, bystrom_exploring_2025, yaaqib_radial_2024, chandra_all-sky_2025}. Clearly, the presence of a $\sim1:10$ satellite has dramatic and wide-spread consequences for its host galaxy. 

The LMC in turn possesses a massive satellite of its own: the SMC. While the SMC-LMC mass ratio is similar to the LMC-MW mass ratio ($\sim1:10$; e.g., \Bref{}), the LMC and SMC exhibit evidence of a much longer interaction history than the LMC and MW. Arguments for the Clouds being a long-lived ($\gtrsim5$ Gyr) binary include (1) a tidal formation channel for the $\sim 150^\circ$ long Stream of HI gas trailing the Clouds requires the SMC to complete multiple orbits about the LMC (e.g., \citealt{besla_simulations_2010}; \Bref{}; \citealt{diaz_constraining_2011, pardy_models_2018, lucchini_magellanic_2020}); (2) the Clouds share similar star formation histories \citep[e.g.,][]{harris_star_2009, weisz_comparing_2013, massana_synchronized_2022}; (3) proper motion measurements of the Clouds indicate they must have had at least one recent close encounter \citep{ruzicka_rotation_2010, zivick_proper_2018}; and (4) it is more cosmologically likely that the Clouds came into the MW as a group rather than meeting for the first time in the MW's outskirts \citep[e.g.,][]{gardiner_numerical_1994, donghia_small_2008, boylan-kolchin_dynamics_2011, busha_statistics_2011}. 

A crucial missing piece of this picture, however, is a framework for understanding the Clouds' impact on each other's DM halos. It is clear from models of the MW-LMC system that the LMC and SMC halos should be distorted in response to their interaction history. These distortions will induce perturbations in the gravitational potential relative to spherically symmetric models. Many authors have included live DM halos in their \textit{N}-body models of the Clouds, both in tailored simulations (e.g. \citealt{besla_simulations_2010}; \Bref{}; \citealt{pardy_models_2018, tepper-garcia_magellanic_2019, lucchini_magellanic_2021, jimenez-arranz_kratos_2024}) and in LMC-SMC analogs found in cosmological simulations \citep[e.g.,][]{chisholm_preinfall_2025}. While these works therefore account for time-dependent perturbations in the Clouds' DM halos, the magnitude of these perturbations, how they evolve over multiple SMC orbits about the LMC, and their effect on the combined gravitational field of the Clouds have not yet been quantified.

Here, we use basis function expansions (BFEs) of a state-of-the-art \textit{N}-body simulation of the Clouds interacting in isolation to identify, quantify, and characterize mutually induced perturbations in the Clouds' DM halos. 

BFEs decompose the density field of a DM halo into a series of biorthogonal basis functions \citep[e.g.,][]{hernquist_self-consistent_1992, weinberg_adaptive_1999, lowing_halo_2011, lilley_general_2023}, facilitating a detailed analysis of deformations to the halo by studying the temporal evolution of the expansion coefficients. This technique has been used with great success to model perturbations to the MW halo in response to the LMC in both idealized simulations \citep[e.g.,][M. S. Petersen et al. in preparation]{garavito-camargo_quantifying_2021, lilleengen_effect_2023, vasiliev_effect_2023} and cosmological analogs \citep[][]{arora_shaping_2025, darragh-ford_shaping_2025}. Here, we apply BFE methods  to the LMC-SMC system for the first time, extending these MW-LMC studies to multiple-pericenter orbits and lower host-mass scales.

We stress that our goal is \textit{not} to exactly reproduce the observed state of the Clouds at present-day. Instead, our simulation treats the Clouds as an isolated binary and does not include the MW. We stop our simulation at a point closely approximating the infall of the Clouds to the MW in Model 2 of \Bref{} to (1) isolate the effect of the Clouds on each other (without needing to disentangle the contributions of the MW's tides); (2) inform us about the state of the LMC-SMC system prior to their infall to the MW; and (3) facilitate comparisons to other 1:10 satellite-host pairs. 

When mapping the perturbations in the Clouds' halos, our goals are (1) to assess the degree to which static, spherical, analytic approximations are appropriate for modeling the LMC-SMC system; (2) to develop a framework for quantifying the perturbations in the LMC halo induced over multiple orbits by the SMC; and (3) to move toward a framework for disentangling the impact of asymmetries in the LMC's halo vs. torques from the SMC in shaping the unusual morphology of the LMC's disk. 

This paper is organized as follows. In Section \ref{sec:meth}, we outline the details of our simulations and the BFEs we use to analyze the LMC and SMC DM halos. Section \ref{sec:bfe} presents our results, describing the DM distributions of the LMC, SMC, and the combined system. We discuss the implications of our findings in Section \ref{sec:disc} before summarizing and offering concluding remarks in Section \ref{sec:conclusions}.

\section{Methods}\label{sec:meth}

Here, we outline the methods used in this work, including a description of our \textit{N}-body simulation of the Clouds (Section \ref{subsec:sims}), an overview of our BFE methods (Section \ref{subsec:exp}), and details on the construction of our LMC and SMC expansions (Sections \ref{subsec:truncation} and \ref{subsec:expansions}).

\subsection{\textit{N}-body Simulation of LMC-SMC Interactions}\label{subsec:sims}

\begin{table}[]
    \centering
    \caption{Initial Halo and Disk Properties of the LMC-SMC \textit{N}-body Simulation}
    \begin{tabular}{c c c}
        \hline
        \hline
        Property & LMC & SMC\\
        \hline
        Halo Profile & Hernquist & Hernquist \\
        Halo Mass [M$_\odot$] & $1.76 \times 10^{11}$ & $1.995 \times 10^{10}$ \\
        Halo Scale Radius [kpc] & 21.4 & 7.3 \\
        Halo Number of Particles & $4.4 \times 10^6$ & $5.0 \times 10^5$ \\ 
        Halo Mass Resolution [M$_\odot$] & $4 \times 10^{4}$ & $4 \times 10^{4}$ \\
        Halo Softening Length [kpc] & 0.08 & 0.08 \\
        Halo $\beta$ (velocity anisotropy) & 0 & 0\\
        Disk Profile & Exponential & Exponential \\
        Disk Mass [M$_\odot$] & $3.6 \times 10^9$ & $5.22 \times 10^8$ \\
        Disk Scale Length [kpc] & 1.7 & 1.1 \\
        Disk Scale Height [kpc] & 0.34 & 0.22 \\
        Disk Number of Particles & $7.2 \times 10^6$ & $1.044 \times 10^6$\\
        Disk Mass Resolution [M$_\odot$] & 500 & 500 \\
        Disk Softening Length [kpc] & 0.05 & 0.05\\
        \hline
    \end{tabular}
    \tablenotetext{}{NOTE - The table columns contain, from left to right, the property, the value for the LMC, and the value for the SMC.}
    \label{tab:sim_params}
\end{table}

\begin{figure}
    \centering
    \includegraphics[width=\columnwidth]{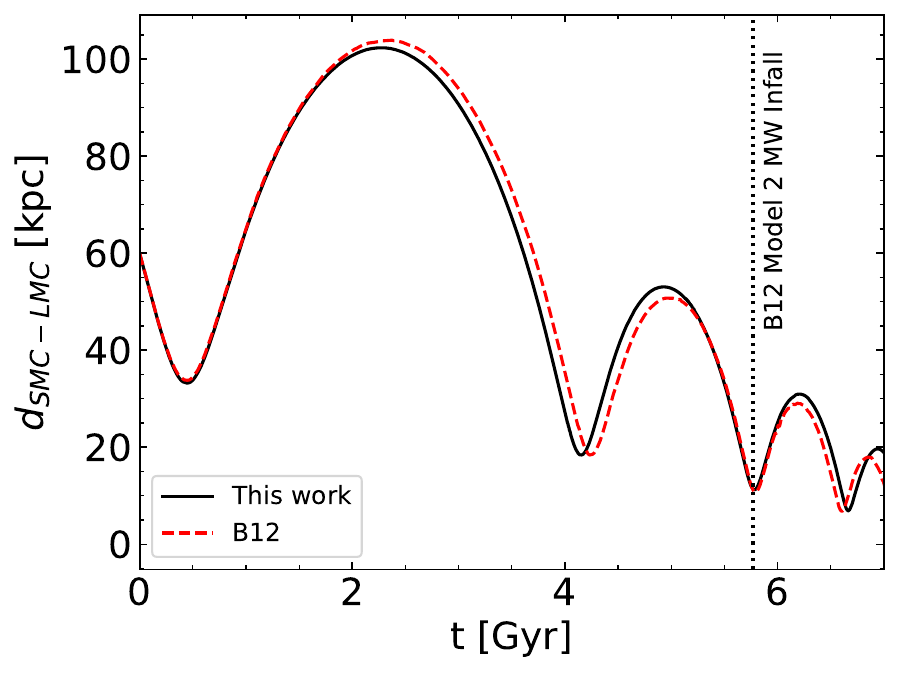}
    \caption{Comparison of the SMC's orbit about the LMC in the simulation by \Bref{} (dashed line) vs. the simulation used in this work (H. Rathore et al. in prep; solid line). Orbits are plotted as the distance between the LMC and SMC centers ($d_{SMC-LMC}$) as a function of simulation time. The latest time considered in this work is $t=5.77$ Gyr (dotted vertical line), which corresponds to the time the LMC and SMC are introduced into the MW in Model 2 of \Bref{}. At this time, the SMC is approaching its third pericenter passage with the LMC.}
    \label{fig:orbit_comp}
\end{figure}

In this section, we describe one simulation from the \texttt{MEGHA} suite of \textit{N}-body simulations of the Clouds (H. Rathore et al. in preparation), which we use in this work. This simulation is an isolated LMC-SMC simulation, describing the Clouds' interaction without the MW, and is motivated by the simulations of the MW-LMC-SMC interaction history presented in \Bref{}.

The initial conditions (ICs) for the LMC and SMC were generated using the code \texttt{MakeGalaxy} (a proprietary version of the publicly available code \texttt{MakeNewDisk}), which approximates the phase-space distribution function using moments of the collisionless Boltzmann equation \citep{Hernquist1993} to generate self-consistent realizations of multiple galaxy components (like the DM halo and stellar disk) under their mutual gravitational influence. 

Table \ref{tab:sim_params} summarizes the initial properties of the DM halos and disks in our simulation. The LMC and SMC have total initial masses of $1.8 \times 10^{11}$ M$_\odot$ and $2.05 \times 10^{10}$ M$_\odot$ respectively. Their live DM halos are initialized with \citet{hernquist_analytical_1990} profiles, with halo masses of $1.76 \times 10^{11}$ M$_\odot$ and $1.995 \times 10^{10}$ M$_\odot$ respectively, and Hernquist scale radii of 21.4 and 7.3 kpc respectively. The above halo parameters of the LMC and SMC are identical to those adopted in \Bref{}. The halo mass resolution of the LMC and SMC is $4 \times 10^{4}$ M$_\odot$ per particle \citep[same as][]{garavito-camargo_hunting_2019}, which is a significant advancement over \Bref{} (DM mass resolution of $\approx 10^6$ M$_\odot$ per particle). The DM particle softening length is 0.08 kpc again following \citet{garavito-camargo_hunting_2019}. The halos are realized with zero velocity anisotropy ($\beta = 0$). 

Our simulation also includes live (\textit{N}-body) stellar disks in both galaxies, which are initialized with exponential density profiles (see Table \ref{tab:sim_params}). The DM halos of the LMC and SMC that we analyze in this paper are therefore self-consistent with the potentials generated by the disks. We defer a detailed description and analysis of the stellar disks to a companion paper (H. Rathore et al. in preparation), and focus on the DM halos for the remainder of this work.

The LMC's halo mass and scale radius are in reasonable agreement with the observational constraints on the LMC's dynamical mass \citep[e.g.,][]{Penarrubia2016, erkal_total_2019, Kacharov2024, Watkins2024}. The SMC is currently in a state of disequilibrium, making dynamical mass estimates unreliable \citep[e.g.,][]{rathore_response_2025}. Given the SMC's stellar mass of $3.1\times10^8 \, \rm{M}_\odot$ \citep[][]{harris_star_2004, stanimirovic_new_2004, skibba_spatial_2012}, abundance-matching gives a DM mass for the SMC of $6\times10^{10} \, \rm{M}_\odot$ at $z=0$ \citep{moster_galactic_2013}, under the assumption that it is an isolated central galaxy. However, the SMC's DM distribution must have been tidally truncated by the LMC. For this simulation, we follow \Bref{} and use a DM mass of $\sim 2 \times 10^{10}$ M$_\odot$, which is a factor of 3 lower than the abundance-matching relation for an isolated central to account for this mass loss. Given the chosen SMC halo mass, the SMC:LMC system is a 1:10 mass ratio encounter (\Bref{}). 

We have checked that the initial halos of the LMC and SMC are stable by evolving each galaxy IC by itself (without the other Cloud) for 7 Gyr and verified that they do not evolve over this timescale. In particular, the halo density profile, velocity dispersion profile, and velocity anisotropy profile remain consistent with the ICs within $10\%$. Further, we have performed convergence tests to ensure that the aforementioned halo properties are not affected by the halo resolution. After verifying the stability of the ICs, we combine the LMC and SMC models using the publicly available \texttt{combine\_ics} code\footnote{https://github.com/himanshrathore/combine\_ics}.

The combined LMC + SMC model is evolved in isolation (without the MW) using the \texttt{GADGET-4} \citep{springel_simulating_2021} \textit{N}-body code\footnote{https://wwwmpa.mpa-garching.mpg.de/gadget4/}. Table \ref{tab:init_phase} shows the position and velocity vector of the SMC with respect to the LMC, in the orbital plane. The simulation was run for a total of 9.7 Gyr, and snapshots were obtained at a cadence of 4.85 Myr. The LMC and SMC completely merge at the end of this simulation.

\begin{table}[]
    \centering
    \caption{Initial position and velocity vector of the SMC with respect to the LMC in our isolated LMC-SMC \textit{N}-body simulation.}
    \begin{tabular}{c c}
        \hline
        \hline
        Quantity & Value \\
        \hline
        $x$ [kpc] & 39.64 \\
        $y$ [kpc] & 44.86 \\
        $z$ [kpc] & 0 \\
        $v_x$ [km s$^{-1}$] & -121.56 \\
        $v_y$ [km s$^{-1}$] & -0.97 \\
        $v_z$ [km s$^{-1}$] & 0 \\
        \hline
    \end{tabular}
    \tablenotetext{}{NOTE - The positions and velocities are specified in the SMC's orbital plane about the LMC.}
    \label{tab:init_phase}
\end{table}

Our isolated LMC-SMC orbit is designed to closely match the pre-MW infall part of the \Bref{} Model 2 orbit (see Figure \ref{fig:orbit_comp}). The center positions of both galaxies at each snapshot are computed from the DM particles only, using the iterative shrinking-sphere algorithm described in \citet{power_inner_2003}. 
This orbit is just one possible realization of the LMC-SMC interaction history, as several other works have presented simulations of the Clouds' interactions \citep[e.g.,][]{besla_simulations_2010, diaz_constraining_2011, pardy_models_2018, lucchini_magellanic_2021, jimenez-arranz_kratos_2024}. 

We choose the \Bref{} isolated SMC-LMC orbit for our simulation because the SMC's pericentric distances and orbit decay rate about the LMC are designed to explain the Clouds' observed, mutually elevated star-formation rates over the past $\sim$3.5 Gyr \citep{harris_star_2009, weisz_comparing_2013, massana_synchronized_2022}. \citet{stierwalt_tiny_2015} reported that isolated pairs of dwarf galaxies with separations of <50 kpc have star formation rates $\geq2.3$ times higher than isolated single dwarf galaxies. The \Bref{} SMC orbit remains at separations $\lesssim50$ kpc from the LMC during the past 3.5 Gyr. Additionally, \Bref{} argue that, with a massive LMC (infall mass $\sim 10^{11}$ M$_\odot$), an \textit{N}-body, decaying orbit for the Clouds must have a high eccentricity ($e\sim0.7$). If the eccentricity is too low, the Clouds will merge rapidly, preventing the observed sustained period of elevated star formation. 

Aside from the orbit, the \Bref{} simulations have also demonstrated reasonable agreement with a variety of external and internal observed features of the Clouds. External features include the SMC's trailing gas Stream \citep{Mathewson1974, Braun2004, Nidever2010} and its stellar counterpart \citep{chandra_discovery_2023, zaritsky_untangling_2025}, the LMC-SMC bridge \citep{Kerr1957, Putman2003, Bruns2005, Zivick2019} and the LMC's present-day Galactocentric position \citep{Kallivayalil2006a, Kallivayalil2006b}. Internal features include: the morphology of the LMC's outer disk \citep{besla_low_2016}, the internal kinematics of the LMC's disk \citep{choi_recent_2022} and the structure and kinematics of the LMC's bar \citep{rathore_precise_2025, rathore_response_2025}. 

Perhaps the biggest concern with the \Bref{} simulations, which would have been a significant impediment to our study \citep[see][]{weinberg_dynamics_1998}, is their low halo resolution ($\approx 10^6$ M$_\odot$). We have addressed this by significantly enhancing the halo resolution in our simulation (Table \ref{tab:sim_params}). 

In \Bref{} Model 2, the Clouds were introduced into the MW's halo at $t=5.77$ Gyr, just before the SMC's third pericenter about the LMC and approximately 1 Gyr before the present-day (see Figure \ref{fig:orbit_comp}). In general, the time of MW infall depends on the specific MW model adopted. However, all models of the MW-LMC interaction agree that the LMC was $\sim200$ kpc from the MW (where the MW's tides are negligible) 1 Gyr ago \citep[see][and references therein]{vasiliev_effect_2023}. As such, we refer to $t=5.77$ Gyr in our simulation as ``MW Infall'' and utilize snapshots before this time to understand the DM distribution of the Clouds before they are impacted by the MW's tides. 

\subsection{Basis Function Expansions with EXP}\label{subsec:exp}

To quantify the time-evolution of the LMC and SMC DM halo shapes during the simulation, we expand the halos in a series of biorthogonal basis functions in spherical coordinates using the the \texttt{EXP} code\footnote{https://github.com/EXP-code} \citep{petersen_exp_2025}. Using the notation of \citet{arora_shaping_2025}, the time-dependent density and potential fields of a DM halo are estimated by

\begin{equation}\label{eqn:density}
    \rho(r, \theta, \phi, t) \approx \sum_{n=0}^{n_{max}}\sum_{l=0}^{l_{max}}\sum_{m=-l}^{l} C_{nlm}(t)\varrho_n(r)Y_l^m(\theta, \phi)
\end{equation}
\begin{equation}\label{eqn:potential}
    \phi(r, \theta, \phi, t) \approx \sum_{n=0}^{n_{max}}\sum_{l=0}^{l_{max}}\sum_{m=-l}^{l} C_{nlm}(t)\psi_n(r)Y_l^m(\theta, \phi)
\end{equation}
\noindent where $n$ denotes the radial order, $l,m$ denote the harmonic order, $\varrho_n$ and $\psi_n$ are the radial basis functions, $Y_l^m(\theta, \phi)$ are the spherical harmonics, and $C_{nlm}$ are time-dependent coefficients. An infinite number of terms would exactly represent the true density field. However, for practical/computational reasons, we are restricted to a finite number of terms. As such, we truncate the expansion at a maximum order $(n_{max}, l_{max})$. In Section \ref{subsec:expansions}, we outline a procedure for finding reasonable values of $n_{max}$ and $l_{max}$ for our expansions. 

A major advantage of \texttt{EXP} over other BFE frameworks is its ability to empirically choose appropriate radial basis functions $\varrho_n$ and $\psi_n$ to ensure fast convergence of the expansion. In particular, \texttt{EXP} uses a Sturm-Liouville solver to find a radial basis whose zeroth-order functions $\varrho_0$ and $\psi_0$ are the spherically averaged density and potential profiles of the input halo, respectively. For more details on this process, we refer the reader to \citet{weinberg_adaptive_1999} and \citet{petersen_exp_2022}. Higher-order terms therefore encode only deviations from the spherically averaged profile, ensuring fast convergence. In contrast, if a specific functional form for the radial basis is chosen, any discrepancies between the halo's spherically averaged profile and the basis require a large number of terms to capture. Once chosen, the basis remains static in time, with time-dependence captured by the coefficients only. 

The coefficients ($C_{nlm}$) are calculated using the DM halo particles at one snapshot as

\begin{equation}\label{eqn:coefs}
    C_{nlm} = \frac{1}{N}\sum_{i=1}^Nm_i\psi_{n}(r_i)Y_l^m(\theta_i\,, \phi_i) \,,
\end{equation}

\noindent where $N$ is the number of DM particles, $m_i$ is the mass of each particle, and $(r_i\,, \theta_i\,, \phi_i)$ are the locations of each particle with respect to the expansion center. We center our expansions at the halo centers identified using shrinking-spheres as described in Section \ref{subsec:sims}. Coefficients are calculated at each simulation snapshot to form time series of the coefficient amplitudes that describe the time-evolution of the halo.

\subsection{Truncating the Order of the Expansions}\label{subsec:truncation}

It is important to choose the expansion truncation length ($n_{max}$ and $l_{max}$) carefully. Too few terms will poorly represent variations in the true density field, while too many terms will overfit the discreteness noise inherent in using a finite set of \textit{N}-body particles to sample the density field \citep{weinberg_high-accuracy_1996}. 

To choose $n_{max}$ and $l_{max}$ for each expansion, we minimize the error of the expansion-reconstructed density field with respect to a smoothed density field calculated directly from the simulation particles. We consider two choices of objective function to calculate the error of our expansions: mean integrated square error (MISE) and mean integrated relative square error (MIRSE). 

For a set of $N$ discrete sampling points $\vec{x}_i$, these are given by
\begin{equation} \label{eqn:MISE}
    \mathrm{MISE} = \frac{1}{N}\sum_{i=1}^N \left(\rho_{Exp}(\vec{x}_i) - \rho_{Part}(\vec{x}_i)\right)^2 
\end{equation}

\begin{equation} \label{eqn:MIRSE}
    \mathrm{MIRSE} = \frac{1}{N}\sum_{i=1}^N \left(\frac{\rho_{Exp}(\vec{x}_i) - \rho_{Part}(\vec{x}_i)}{\rho_{Part}(\vec{x}_i)}  \right)^2, 
\end{equation}

\noindent where $\rho_{Exp}(\vec{x}_i)$ is the density derived from the expansion, and $\rho_{Part}(\vec{x}_i)$ is the density inferred directly from the simulation particles by calculating the density of the 1000 closest simulation particles to $\vec{x}_i$. We have verified that this method of calculating the smoothed simulation particle density is capable of recovering the Hernquist density profiles of the ICs to within 5\% within a radius of 150 kpc in both the LMC and SMC.

The difference between MISE and MIRSE lies in their weighting. MISE penalizes absolute errors, so it is more tolerant of large relative errors in the low-density outskirts of a halo in favor of producing a good match to the high-density inner halo. MIRSE, in contrast, penalizes relative errors directly and therefore weights the inner halo and outskirts equally in this sense. In Section \ref{subsec:expansions}, we describe each of our expansions and the rationale behind our choice of objective function for each. 

Figure \ref{fig:MIRSE} summarizes the error analysis for each expansion. In Appendix \ref{apdx:validation}, we validate our choices of truncation order for each expansion. 

\subsection{Expanding the LMC and SMC Dark Matter Halos}\label{subsec:expansions}

\begin{table*}[]
    \centering
    \caption{Summary of the LMC and SMC BFEs}
    \begin{tabular}{c c c c c}
        \hline
        \hline
        Expansion & Components & Objective Function & 
        $n_{max}$ & $l_{max}$ \\
        \hline
        LMC & All LMC particles & MIRSE & 17 & 10 \\
        SMC & All SMC particles & MIRSE & 17 & 10 \\
        Bound SMC & Bound SMC particles with $r_{SMC} < 50$ kpc & 
        MISE & 12 & 5 \\
        \hline
    \end{tabular}
    \tablenotetext{}{NOTE - The table columns contain, from left to right, the name of each expansion, the particles used to construct the expansion, the objective function used in the error-minimization procedure to choose the truncation order, and the truncation order in $n$ and $l$. Radial basis functions for each expansion match the Hernquist profiles given in Table \ref{tab:sim_params}.}
    \label{tab:expansions}
\end{table*}

\begin{figure*}
    \centering
    \includegraphics[height=6.6cm]{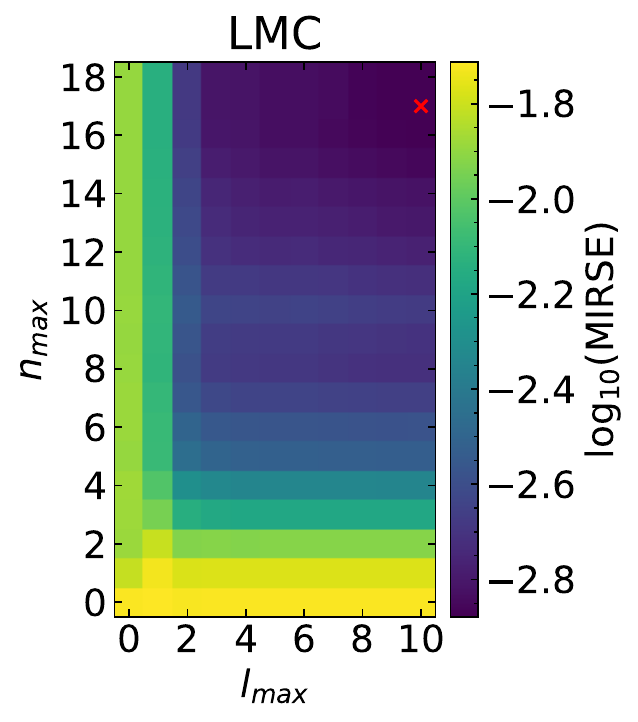}
    \includegraphics[height=6.6cm]{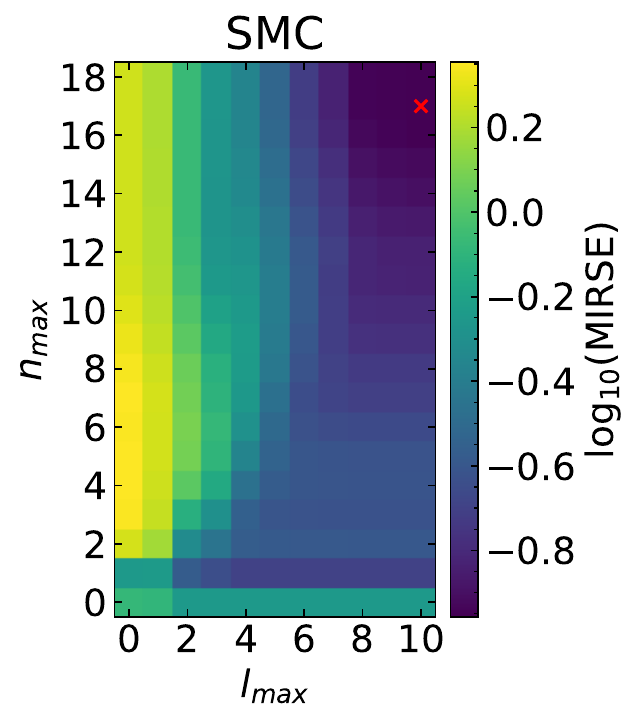}
    \includegraphics[height=6.6cm]{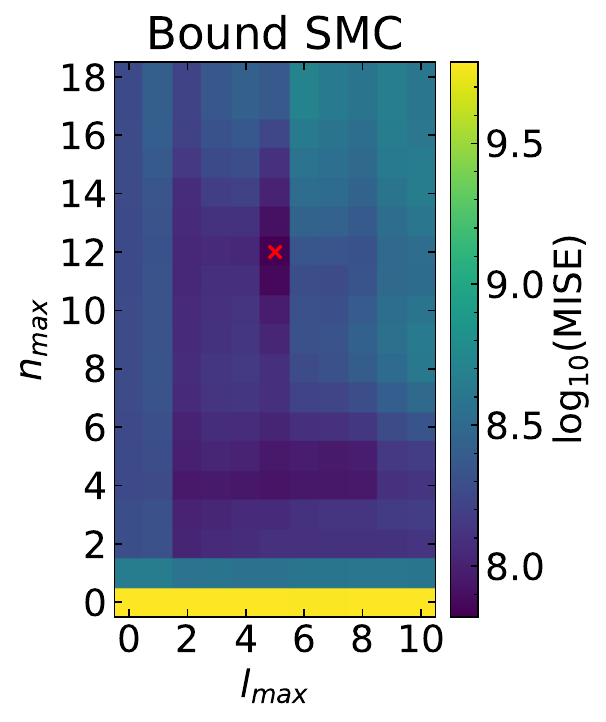}

    \caption{Error in each expansion (LMC, SMC, Bound SMC) with respect to the particle density as a function of the truncation order, $n_{max}$ and $l_{max}$, calculated at the MW infall snapshot. A red x shows the value of $n_{max}$ and $l_{max}$ where the error is minimized (see Table \ref{tab:expansions}). \textit{Left Panel:} The LMC Expansion is truncated according to MIRSE (Equation \ref{eqn:MIRSE}) at $(n_{max},  l_{max}) = (17, 10)$. \textit{Center Panel:} The SMC Expansion is truncated at $(n_{max},  l_{max}) = (17, 10)$ according to an MIRSE minimization. \textit{Right Panel:} The Bound SMC Expansion is constructed to minimize MISE (Equation \ref{eqn:MISE}) to prioritize the reconstruction of the high-density inner halo, resulting in a truncation order of $(n_{max},  l_{max}) = (12, 5)$.}
    \label{fig:MIRSE}
\end{figure*}

We construct three different BFEs, one for the LMC (LMC Expansion) and two for the SMC. The two expansions for the SMC use either all the DM particles (SMC Expansion) or only bound particles (Bound SMC Expansion). The radial bases are generated using the analytic Hernquist density profiles adopted for the halo ICs, i.e. the zeroth-order radial basis functions match the DM halo profiles specified in Table \ref{tab:sim_params}. Our expansions are summarized in Table \ref{tab:expansions}, illustrated in Figure \ref{fig:composite}, and described as follows:

\textit{LMC Expansion}: The LMC halo expansion (panel (\textit{c}) of Figure \ref{fig:composite}) is calculated using all DM particles that were originally assigned to the LMC. This expansion captures the evolution of the entirety of the LMC's halo throughout its interaction with the SMC. 

As we are interested in perturbations induced in the LMC from the SMC at large radii, we use MIRSE to choose the truncation order, ensuring the expansion captures features in the outskirts of the halo. The left panel of Figure \ref{fig:MIRSE} shows the MIRSE for different choices of truncation order in the LMC expansion, calculated at the MW infall snapshot (after $t=5.77$ Gyr of evolution). For the LMC, we calculate the MIRSE with $100^3$ sampling points uniformly distributed within a cube of 200 kpc side length centered on the expansion center. The MIRSE is minimized for $n_{max}=17$ and $l_{max}=10$, so we choose these values to truncate the LMC expansion.

\textit{SMC Expansion}: The expansion of the SMC's halo is calculated using all DM particles that were originally assigned to the SMC (panel (\textit{d}) of Figure \ref{fig:composite}). This expansion tracks the evolution of the SMC's DM debris as it is tidally stripped by the LMC. It is also used to reproduce the combined DM distribution of the system when added to the LMC expansion (panel (\textit{a}) of Figure \ref{fig:composite}). 

We are interested in accurately capturing the outskirts/debris of the SMC and therefore choose MIRSE to truncate the SMC expansion. The center panel of Figure \ref{fig:MIRSE} shows the MIRSE analysis for the SMC expansion. We take $50^3$ samples within a 100 kpc sided cube such that the sampling density is equal to that for the LMC expansion. A larger number of terms compared to the LMC are needed to capture the structure of the SMC's debris, i.e. expansions up to $n_{max}$ of 16 or $l_{max}$ of 8 still produce nonnegligible gains in MIRSE. Like the LMC, the MIRSE for the SMC+debris expansion is minimized for $n_{max}=17$ and $l_{max}=10$.

\textit{Bound SMC Expansion}: To track the SMC's mass loss and quantify the shape of its inner halo ($\lesssim$ 10 kpc) throughout the simulation, we construct an alternative SMC expansion using only bound DM particles within 50 kpc of the SMC's density cusp (panel (\textit{e}) of Figure \ref{fig:composite}). At each time step, we use the BFE potential calculated from all SMC DM particles (the SMC Expansion) to calculate the total energy of each SMC DM particle $E_{tot} = v_{SMC}^2/2 - \phi_{SMC}$, where $v_{SMC}$ is the DM particle's velocity with respect to the SMC center. We use only DM particles with $E_{tot}<0$ and positions within 50 kpc of the SMC's center to construct the Bound SMC Expansion. 

Note that, while our simulations include a stellar disk (and therefore, the DM reacts to the disk's presence), we do not include the disk's contribution to the potential when computing bound particles. We have checked that including the disk impacts the inferred bound mass of the SMC by < 5\% at the MW infall snapshot, validating this simplification. Expansions of the LMC and SMC disks are beyond the scope of this work and will instead be presented in an upcoming companion paper (H. Rathore et al. in preparation).

For the Bound SMC Expansion, we are primarily interested in accurately capturing the shape of the inner halo. We choose MISE to truncate the Bound SMC expansion to ensure the highly distorted inner halo is well captured. The right panel of Figure \ref{fig:MIRSE} shows the MISE minimization for this expansion, which uses the same 100 kpc wide sampling grid as the SMC Expansion. The MISE is minimized for $n_{max}=12$ and $l_{max}=5$, at which we truncate the Bound SMC Expansion.

\section{Reconstruction of the LMC-SMC Dark Matter Distribution}\label{sec:bfe}

\begin{figure*}
    \centering
    \includegraphics[width=\textwidth]{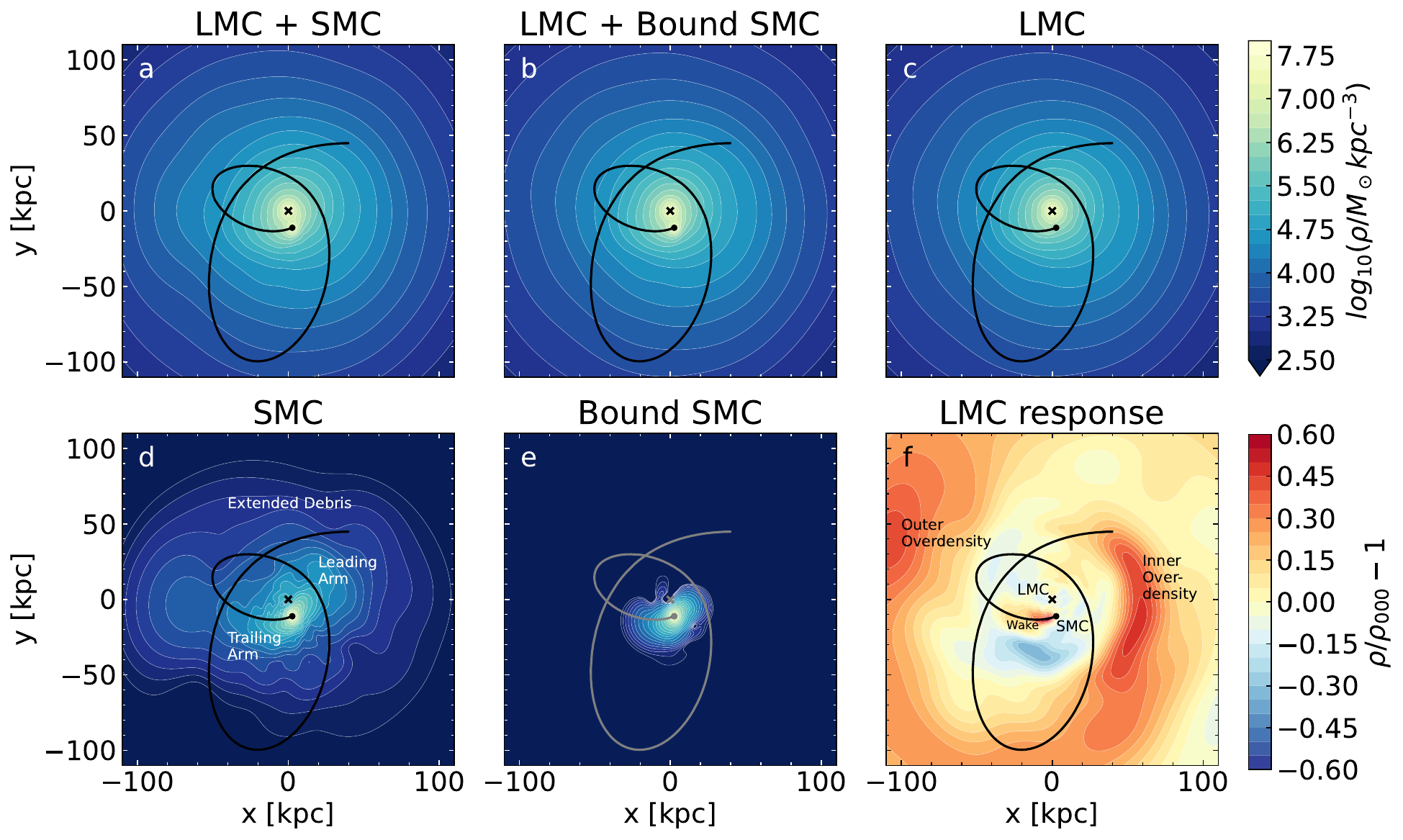}
    \caption{BFE-reconstructed density fields of the isolated LMC-SMC system (no MW) at MW infall. Density fields are plotted in the SMC's orbital plane about the LMC. 
    The locations of the LMC and SMC centers are marked with an x and dot, respectively. The SMC's past orbit is marked with a solid line. Panels (\textit{a})-(\textit{e}) share the same contour scale. 
    (a) The density field of the entire system (\textit{c} + \textit{d}). The DM distribution of the LMC-SMC system is distorted at MW infall, exhibiting nonspherical isodensity contours. 
    (b) The sum of the LMC Expansion and Bound SMC Expansion (\textit{c} + \textit{e}). The asymmetry in the combined DM distribution at negative \textit{x} is due to the SMC's extended debris field (see panel (\textit{d})).  
    (c) The LMC Expansion. The LMC halo deviates from sphericity due to perturbations from the SMC (see panel (\textit{f})).  
    (d) The SMC Expansion. The SMC's debris extends to $\sim100$ kpc from the LMC's center.
    (e) The Bound SMC Expansion. The SMC's halo is tidally elongated by the LMC over multiple pericenters.
    (f) The density contrast of the LMC Expansion (panel (\textit{c})) computed with respect to the $n=0$, $l=0$ term (see equation \ref{eqn:odens}). Perturbations to the LMC's halo induced by the SMC include a dynamical friction wake and two overdensities (at $\sim$60 and $\sim$100 kpc). The overdensities result from the LMC's density center displacement after the SMC's second and third pericenter passages. This figure is comparable to Figure 1 of \citealt{garavito-camargo_quantifying_2021}) for the LMC-MW system. The SMC's wake at MW infall is roughly twice as strong (peak contrast of $\sim 0.4$) as the LMC's wake in an isotropic MW at present-day ($\sim$0.2; \citealt{garavito-camargo_hunting_2019}).}
    \label{fig:composite}
\end{figure*}

In this section, we present and analyze the BFEs of the LMC (Section \ref{subsec:LMC_exp}), SMC (Section \ref{subsec:debris_exp}), and Bound SMC (Section \ref{subsec:SMC_exp}), before discussing the combined DM distribution of the LMC-SMC system in Section \ref{subsec:comb_exp}. The reconstructed DM density fields for each of these expansions are summarized in Figure \ref{fig:composite}.

\subsection{The LMC's Dark Matter Halo}\label{subsec:LMC_exp}

\begin{figure*}
    \centering
    \includegraphics[width=\textwidth]{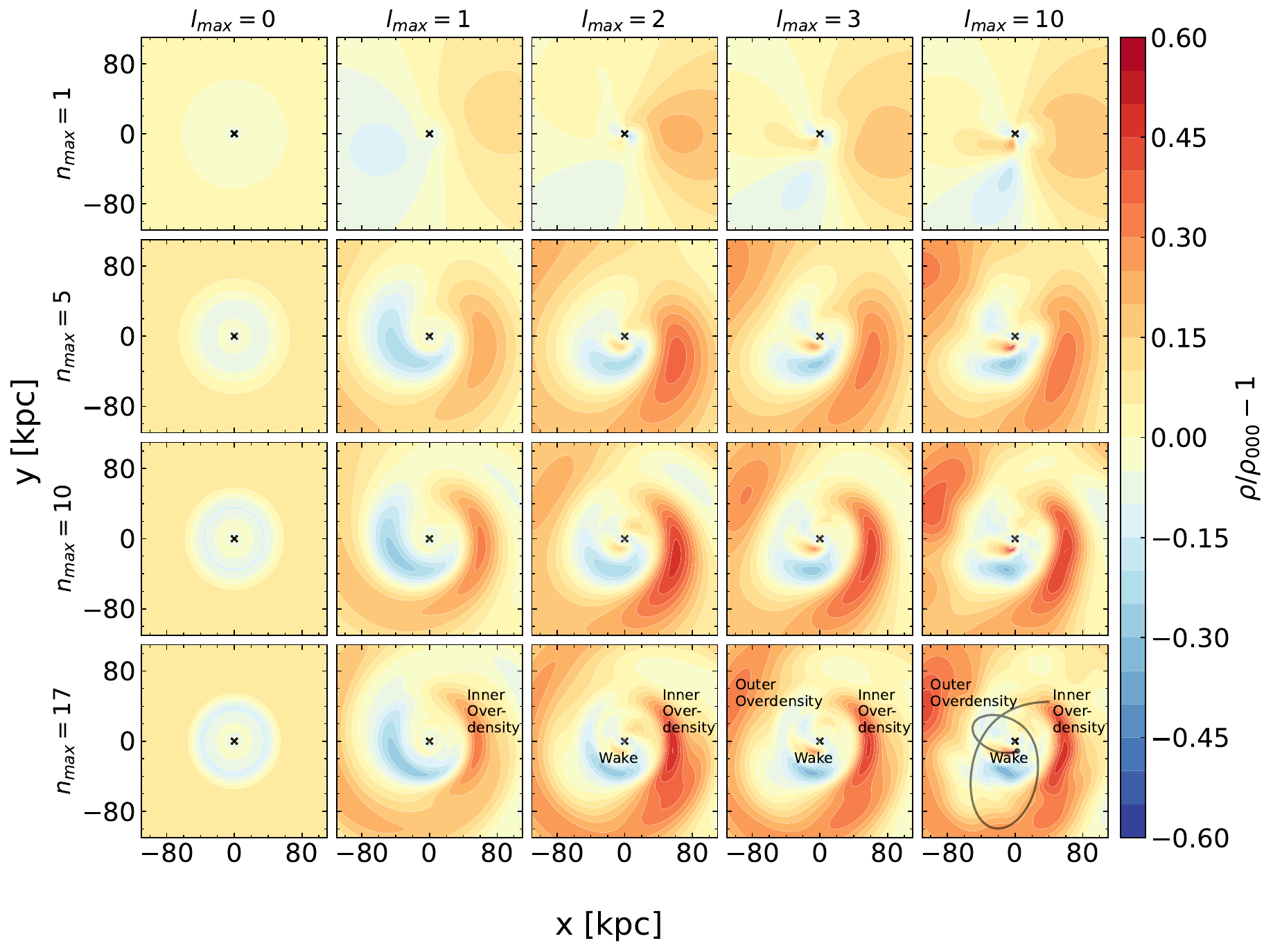}
    \caption{Density contrast (see Equation \ref{eqn:odens}) of the LMC Expansion (no SMC expansion is included) at MW infall for different choices of the maximum expansion order in radial ($n_{max}$; varies between rows) and harmonic ($l_{max}$; varies between columns) terms. The LMC's center is marked with an `x.' Increasing the radial (harmonic) order of the expansion captures more radial (azimuthal) variations in the density field. The density center displacement from the SMC's third pericenter passage (Inner Overdensity) is captured by the $l=1$ terms. Higher harmonics ($l\geq2$) capture the Outer Overdensity from the density center displacement during the second SMC pericenter and the SMC's wake. Truncating the expansion at $n_{max}=17$ and $l_{max}=10$ accurately recovers the peak amplitudes of the perturbations (see Figure \ref{fig:MIRSE}); this is the fiducial expansion order for the LMC in this study. 
    }
    \label{fig:odens_orders}
\end{figure*}

\begin{figure}
    \centering
    \includegraphics[width=\columnwidth]{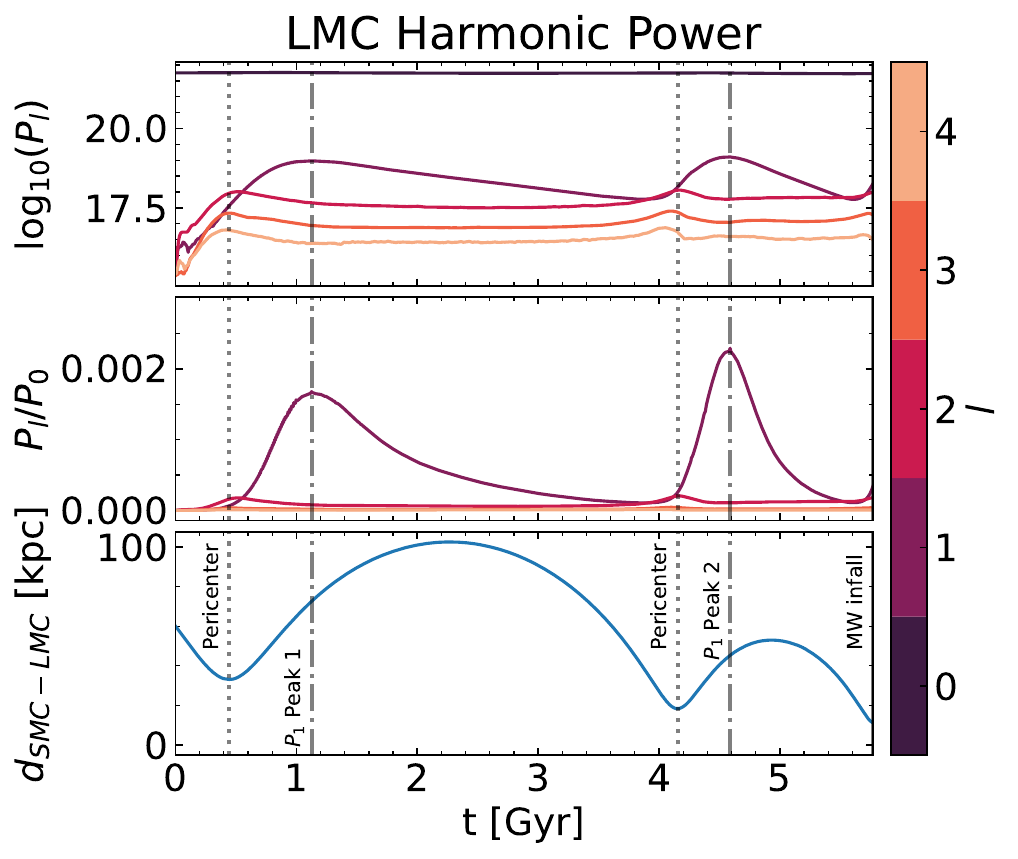}
    \caption{Gravitational power ($P_l$; equation \ref{eqn:power}) in the first five harmonic orders of the LMC Expansion (no SMC) as a function of time. 
    {\it Top panel:} the log power in each harmonic. {\it Center panel:} 
    the power in each harmonic normalized by the power in the monopole terms ($l=0$). {\it Bottom panel:} the distance between the LMC and SMC centers ($d_{SMC-LMC}$) as a function of time, with pericenter times (at $t=0.44$ and $4.15$ Gyr) marked by dotted lines. MW infall is at the right edge of the plot. The SMC induces a spike in the LMC's dipole ($l=1$) power {\it after} each pericenter (marked by dot-dashed lines), which is a characteristic signature of the motion of the inner LMC halo owing to the displacement of the LMC density center. Following the first (second) peak, the dipole power decays exponentially with a decay timescale of 920 (300) Myr, before being reexcited by the SMC during a subsequent pericenter passage. In contrast, quadrupole and higher terms ($l\geq2$), which build up the SMC's wake (see Figure \ref{fig:odens_orders}), peak at each SMC pericenter. These $l\geq2$ terms remain elevated compared to the start of the simulation, indicating that the wake persists throughout, despite its variations in strength tied to the SMC's orbit.}
    \label{fig:modes}
\end{figure}

\begin{figure*}
    \centering
    \includegraphics[width=\textwidth]{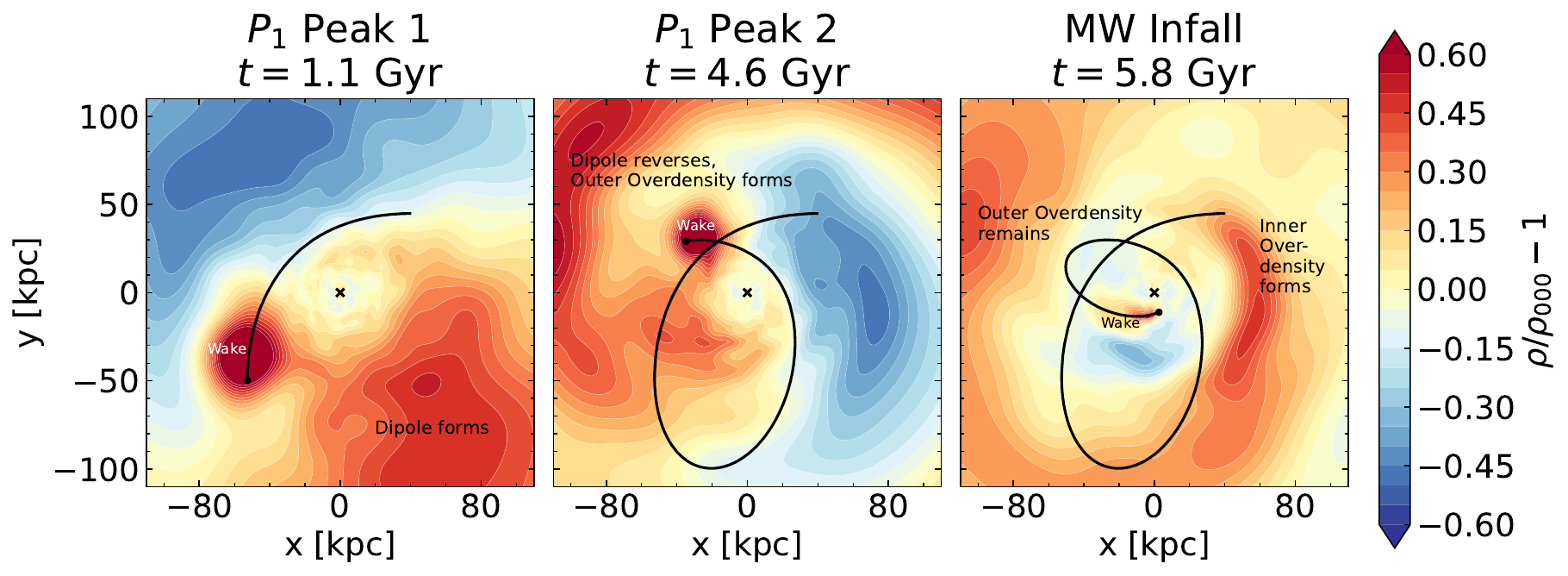}
    \caption{Formation of the perturbations in the LMC halo. Panels show the density contrast (Equation \ref{eqn:odens}) of the LMC Expansion in the SMC orbital plane at the time of each $l=1$ power peak identified in Figure \ref{fig:modes} and at MW infall (right panel is identical to panel (\textit{f}) of Figure \ref{fig:composite}). \textit{Left panel}: the dipole associated with the SMC's first pericentric passage is at its peak power. The SMC's wake is also strong at this time. \textit{Center panel}: time of peak dipole power following the SMC's second pericenter. By this time, the density dipole in the LMC halo has reversed phase. \textit{Right panel}: by the time of MW infall, the LMC's density field is a superposition of the wake, Outer Overdensity (left over from the SMC's second pericenter), and the Inner Overdensity (associated with a dipole forming at smaller radii due to the SMC's imminent third pericentric passage).}
    \label{fig:evolution}
\end{figure*}

\begin{figure}
    \centering
    \includegraphics[width=\columnwidth]{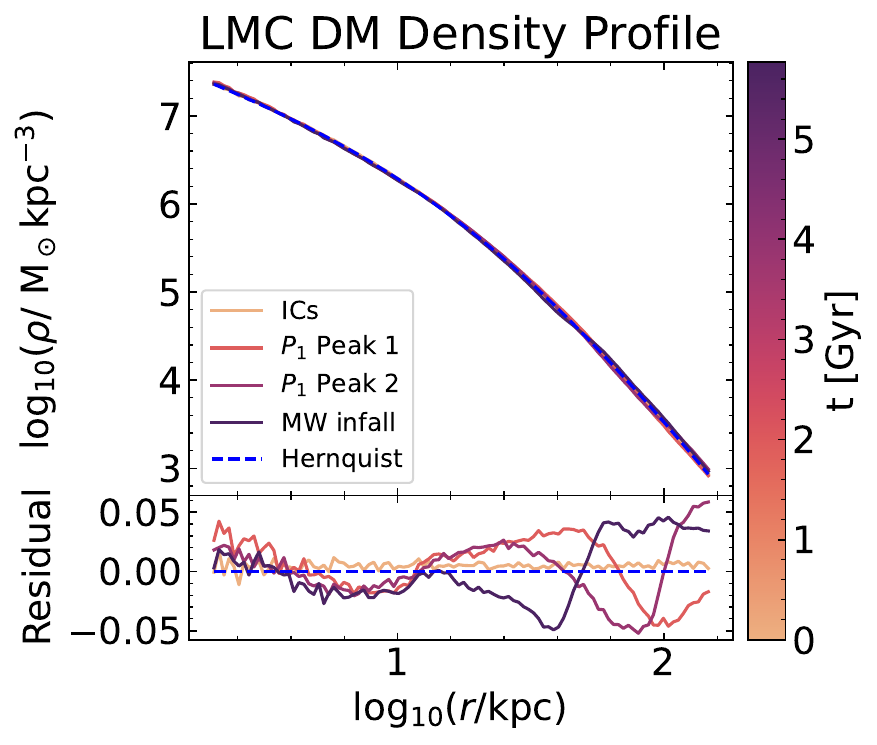}
    \caption{\textit{Top panel:} time-evolution of the LMC's spherically averaged DM density profile measured using simulation particles at the IC, each dipole power ($P_1$) peak identified in Figure \ref{fig:modes}, and at MW infall. The color of each line indicates the simulation time when the profile was measured. The dashed line indicates the analytic Hernquist profile used to generate the LMC IC. \textit{Bottom panel:} ratio of the measured density profiles with respect to the Hernquist IC (Residual = $\mathrm{log}(\rho_{Part}(r)) - \mathrm{log}(\rho_{Hern}(r)) = \mathrm{log}[\rho_{Part}(r)/\rho_{Hern}(r)]$). The LMC's spherically averaged DM density profile shows minimal evolution over the simulation, despite the distortions induced by the SMC. }
    \label{fig:lmc_profiles}
\end{figure}

\begin{figure}
    \centering
    \includegraphics[width=\columnwidth]{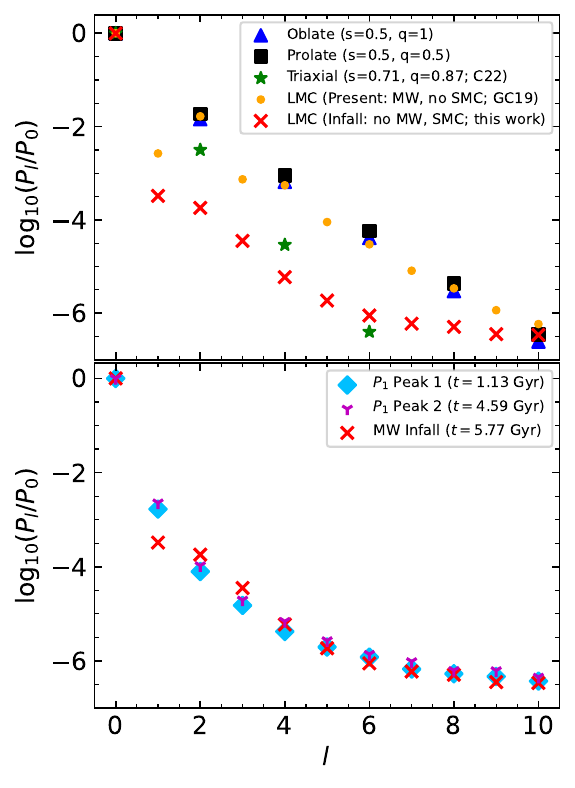}
    \caption{Understanding the LMC's halo shape. \textit{Top panel}: gravitational power of each harmonic in the given expansion, normalized by the monopole ($P_l/P_0$). Red x's denote the LMC Expansion at MW infall. Results for expansions of halos of specified shapes are plotted for comparison 
    (see Table \ref{tab:templates}): oblate; prolate; triaxial (cosmological LMC-mass halos from \citet[][C22]{chua_impact_2022}); and an expansion of the LMC at the present-day snapshot of simulation 3 of \citet[][GC19]{garavito-camargo_hunting_2019}, which has no SMC, but is subjected to MW tides. The LMC Expansion at infall has less power in higher harmonics than the comparison models; it is more spherical. However, the SMC induces asymmetric perturbations -- there is power in odd harmonics, which are absent in symmetric models. The MW induces a quadrupole ($l=2$) that is 2 orders of magnitude stronger than that caused by the SMC. 
    \textit{Bottom panel:} comparison of the LMC Expansion at three times, MW infall (red x's), and the first and second $P_1$ peaks identified in Figure \ref{fig:modes}. The dipole strength varies by an order of magnitude, depending on the SMC's orbit, and can be as strong as the dipole induced by the MW (upper panel). At infall, the $l=2$ and $l=3$ terms are stronger than in the past. 
    }
    \label{fig:template_power_lmc}
\end{figure}

We plot the BFE-reconstructed density field of the LMC DM halo at the time of MW infall in panel (\textit{c}) of Figure \ref{fig:composite}. The isodensity contours reveal that the SMC has induced deviations from sphericity at nearly all radii in the LMC's halo. 

To illustrate the distortions in the LMC's halo induced by the SMC, we plot the density contrast with respect to the monopole in panel (\textit{f}) of Figure \ref{fig:composite}, which is given by

\begin{equation}\label{eqn:odens}
    \delta\rho=\frac{\rho}{\rho_{000}}-1
\end{equation}

\noindent where $\rho$ is the BFE-reconstructed density using all expansion terms, and $\rho_{000}$ is the $n=l=m=0$ term in the density expansion. We identify the major structures in the LMC's halo induced by the SMC as follows:

\textit{The SMC's wake}. The SMC, as an infalling massive satellite, is expected to excite a trailing overdensity in the host LMC halo due to orbital resonances between the SMC and the LMC's DM particles \citep{weinberg_orbital_1986}. This overdensity is identified as the SMC's dynamical friction wake \citep{chandrasekhar_dynamical_1943, mulder_dynamical_1983} and is responsible for the decay of its orbit. The SMC's wake extends for roughly 20 kpc behind the SMC and reaches a peak overdensity of roughly 0.4 at the time of MW infall. This density contrast is roughly twice as strong as the LMC's wake in isotropic models of the MW (contrast of $\sim 0.2$; \citealt{garavito-camargo_hunting_2019}). The SMC orbits the LMC at a lower speed than the LMC orbits the MW, which explains this difference in wake strength. In a more realistic, radially anisotropic host halo, the strength of the wake is expected to increase \citep{garavito-camargo_hunting_2019, rozier_constraining_2022}. 

\textit{DM density dipoles}. In addition to the dynamical friction wake, the other major component of a host halo's response to an infalling satellite is the excitation of a self-gravitating, weakly damped dipole ($l=1$) mode in the host. This mode can be interpreted as the displacement of the host's density center with respect to its outer halo due to the motion of the inner halo about the orbital barycenter \citep{weinberg_self-gravitating_1989}. This phenomenon is sometimes referred to as the ``collective response'' \citep[e.g.,][]{weinberg_self-gravitating_1989, weinberg_nonlocal_1993, gomez_fully_2016, garavito-camargo_hunting_2019, garavito-camargo_quantifying_2021, conroy_all-sky_2021}, although we will follow \citet{darragh-ford_shaping_2025} and refer to it as ``density center displacement'' hereafter. In the host halo's density field, the density center displacement manifests as an overdensity leading the satellite orbit at larger radii and a corresponding underdensity trailing the satellite. 

At the epoch of MW infall, we identify two overdensities arising from successive dipole excitations as the LMC's density center is displaced during the SMC's second and third pericenter passages. The \textit{Inner Overdensity} corresponds to the SMC's third pericenter passage in the LMC halo at MW infall. This overdensity leads the SMC's present-day position, reaching a peak contrast of $\sim 0.5$ at roughly 60 kpc from the LMC's center. The corresponding underdensity lies outside the SMC's wake, reaching a density contrast of $\sim -0.3$. The \textit{Outer Overdensity} in the LMC halo lies at a distance of $\sim 100$ kpc, opposite to the Inner Overdensity. The Outer Overdensity has a peak strength of $\sim 0.4$, and is created by the displacement of the LMC's density center during SMC's previous (second) pericenter at $t=4.15$ Gyr. 

The wake and density center displacement seen here in the LMC are qualitatively similar to the structures predicted in the MW halo induced by the LMC \citep{garavito-camargo_hunting_2019, petersen_reflex_2020, garavito-camargo_quantifying_2021, rozier_constraining_2022, lilleengen_effect_2023, vasiliev_dear_2024} and the Sagittarius dwarf spheroidal \citep[Sgr;][M. S. Petersen et al. in preparation]{laporte_influence_2018}. These results are also consistent with those seen in cosmological simulations with massive satellites \citep[][]{gomez_fully_2016, arora_shaping_2025, darragh-ford_shaping_2025}. Our work here extends these studies to low-mass hosts, providing evidence that these halo distortions are a common consequence of a massive satellite infall.

Figure \ref{fig:odens_orders} shows how different terms in the LMC BFE build up the wake and the Inner and Outer Overdensities. The $l=1$ terms at radial orders $n\leq5$ capture primarily the Inner Overdensity, while adding higher-order $n$ terms begins to capture the Outer Overdensity as well by providing support at additional radial scales. Adding the quadrupole ($l=2$) terms begins to include the SMC's Wake, and refines the structure of the Outer Overdensity. 

The MIRSE analysis concluded that the LMC Expansion should be truncated at $l_{max}=10,\, n_{max}=17$. This corresponds to the bottom right panel, which captures the peak density contrasts and features described above (see Appendix \ref{apdx:validation}). However, we note that a much lower truncation order ($l_{max}=3,\, n_{max}=10$) can still qualitatively describe all three major features (Wake, Inner and Outer Overdensities) seen in the LMC halo. This demonstrates that, while including additional terms up to $l_{max}=10,\, n_{max}=17$ improves the expansion's recovery of the density field (Figure \ref{fig:MIRSE}), the identification of individual perturbations in the LMC halo is robust to significant variations of the truncation order.

To understand the time-evolution of the LMC's halo, we track the gravitational power of the harmonics of the LMC Expansion as a function of time. The gravitational power of a given harmonic is

\begin{equation}\label{eqn:power}
    P_l = -2W_l = \sum_{n=0}^{n_{max}}\sum_{m=0}^l C_{nlm}^2,
\end{equation}

\noindent where $W$ is the gravitational potential energy stored in the harmonic (again using the notation of \citealt{arora_shaping_2025}). Figure \ref{fig:modes} shows the time-evolution of the power in the first five LMC harmonics, along with the SMC's orbit as a timing reference. 

The dipole power ($P_1$), which captures the density center displacement of the LMC, peaks strongly 710 Myr after the SMC's first pericenter passage and 430 Myr after its second pericenter, and decays exponentially following each peak. $P_1$ is starting to grow again at MW infall, as the SMC approaches its third pericenter. This $P_1$ behavior indicates that the LMC halo's dipole response is repeatedly excited by the SMC during each of its pericenter passages, while decaying between these excitations when the SMC is farther away from the LMC's center.

Higher-order harmonics ($P_{l\geq2}$) peak at the SMC's pericenters, indicating that the wake strength increases as the SMC passes through the high-density inner halo of the LMC. However, the strength of the quadrupole power is an order of magnitude weaker than that of the dipole. 

These results -- dipole (quadrupole) power peaks in the host following (during) satellite pericenter passages -- confirm findings from cosmological \citep[][]{gomez_fully_2016, arora_shaping_2025, darragh-ford_shaping_2025} and idealized (\citealt{lilleengen_effect_2023}, M. S. Petersen et al. in preparation) simulations, reinforcing that this host halo behavior is common in $\sim1:10$ mergers.

At our choice for the epoch of MW infall, the strength of the dipole is significantly lower than it was soon after the SMC's second pericenter. To understand the perturbations in the LMC's halo at times of maximal distortion, we compare the density contrast in the LMC halo at MW infall to times when the dipole power reaches a maximum in Figure \ref{fig:evolution}. In the left panel, we can see a clear dipole from the LMC's density center displacement during the SMC's first pericenter passage. Following the SMC's second pericenter (center panel), the LMC's density center is again displaced, exciting a dipole with a roughly opposite phase. The overdensity associated with this second pericenter dipole persists to MW infall (right panel), and is identified as the Outer Overdensity. At MW infall, the density center displacement during the SMC's ongoing third pericenter passage has excited an additional dipole lobe at $\sim60$ kpc, forming the Inner Overdensity. 

Ultimately, we interpret the LMC's halo structure at MW infall as the complex result of entrained dipole excitations at successive SMC pericenters \citep[see, e.g.][for discussions of this mechanism in general systems]{weinberg_weakly_1994, weinberg_new_2023}, plus the SMC's dynamical friction wake. We emphasize that the SMC's wake is comparatively weak at MW infall, as it reached stronger density contrasts in the past. Clearly, the shape of the LMC halo is both complex and time-dependent.

We now turn to characterizing the global structure of the LMC halo in response to the SMC. Figure \ref{fig:lmc_profiles} shows the evolution of the LMC's spherically averaged density profile throughout the simulation, measured from the simulation particles at times corresponding to $P_1$ maxima and MW infall to capture the range of severity of LMC halo distortions. We compare these measured profiles to the Hernquist profile that was used to generate the LMC IC, finding that the LMC's spherically averaged density profile remains nearly static throughout the simulation, despite the asymmetries induced by the SMC.

\begin{table}[]
    \centering
    \caption{Summary of comparison halo shapes}
    \begin{tabular}{c c c}
        \hline
        \hline
        Halo Shape & $s$ & $q$ \\
        \hline
        Oblate & 0.5 & 1 \\
        Prolate & 0.5 & 0.5 \\
        Triaxial \citep{chua_impact_2022} & 0.71 & 0.87 \\
        \hline
    \end{tabular}
    \tablenotetext{}{NOTE - Columns list: the name of the halo shape model, with reference if applicable; the minor-to-major axis ratio $s=c/a$; and the intermediate-to-major axis ratio $q=b/a.$}
    \label{tab:templates}
\end{table}

Inspired by \citet{garavito-camargo_quantifying_2021}, we compare the LMC Expansion to a library of ``template'' halo expansions, summarized in Table \ref{tab:templates}. Each template halo is described by its minor-to-major ($s=c/a$) and intermediate-to-major ($q=b/a$) axis ratio. We choose extreme examples of oblate and prolate halos, as well as a triaxial halo with axis ratios of a typical LMC-mass galaxy in the Illustris-TNG L25n512 simulation with a fiducial feedback model (``Fiducial'' model for $10^{11\mathrm{-}11.5}\,\rm{M}_\odot$ halos in Figure 7 of \citealt{chua_impact_2022}). 

We also perform an expansion of the LMC at the present-day snapshot of \citet{garavito-camargo_hunting_2019}'s Simulation 3 to compare the effect of the MW's tides vs. the SMC on the LMC halo. \citet{garavito-camargo_quantifying_2021} performed an expansion of their (initially spherical) LMC halo in the \citet{garavito-camargo_hunting_2019} simulations, but did not plot it in this fashion. Therefore, we have used their simulation but performed an expansion to ($n_{max}, l_{max})=(17,10)$ according to our methodology described in Sections \ref{subsec:exp} and \ref{subsec:truncation}, using MIRSE to determine the truncation order.

The top panel of Figure \ref{fig:template_power_lmc} shows the results of these comparisons. We plot the gravitational power of each harmonic in each expansion normalized by the strength of the monopole. As discussed in \citet{garavito-camargo_quantifying_2021}, the template oblate, prolate, and triaxial halos lack power in odd harmonics due to their inherent symmetries. The presence of power in odd $l$ terms for the two LMC simulations therefore signifies the presence of the asymmetric perturbations induced by the SMC or the MW.

In the \citet{garavito-camargo_hunting_2019} LMC models (no SMC), the MW's tides have a stronger effect on the LMC's global shape than what we find for the SMC's influence on the LMC at the epoch of MW infall. In particular, the strength of the quadrupole in the present-day LMC is comparable to that of the oblate and prolate halos. This indicates that the MW's tides have elongated the LMC's DM halo significantly \citep[e.g.,][]{garavito-camargo_quantifying_2021, lilleengen_effect_2023, vasiliev_dear_2024}. 

The bottom panel of Figure \ref{fig:template_power_lmc} compares the LMC Expansion at different epochs. The $l=1$ power is weakest at MW infall, but can reach strengths comparable to the impact of the MW on the LMC in \citet[upper panel]{garavito-camargo_hunting_2019}. Furthermore, both the Inner and Outer Overdensities persist at MW infall, resulting in higher power in $l=2,3$ than at earlier times. While we cannot directly extrapolate these results at MW infall to the present-day, it is clear that the SMC's impact on the LMC (in the absence of the MW) is nonnegligible compared to the impact of MW tides on the LMC (in the absence of the SMC; \citealt{garavito-camargo_quantifying_2021}). Therefore, the present-day LMC must be shaped by \textit{both} the SMC and the MW. However, exactly how the LMC halo perturbations identified in this work will evolve further in the presence of the MW is unclear; full MW-LMC-SMC simulations are needed to address this issue.

\subsection{The SMC's Dark Matter Debris}\label{subsec:debris_exp}

\begin{figure*}
    \centering
    \includegraphics[height=2.75in]{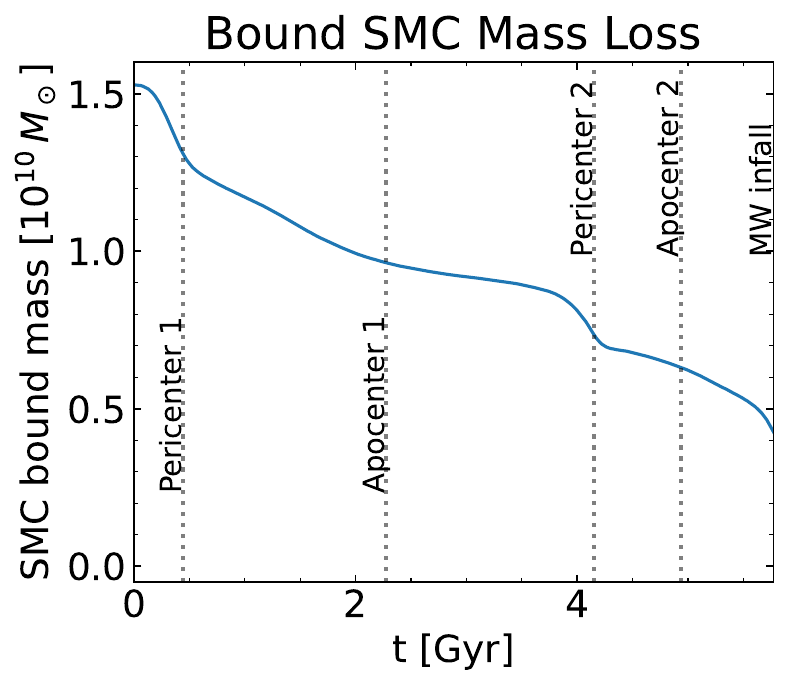}
    \includegraphics[height=2.75in]{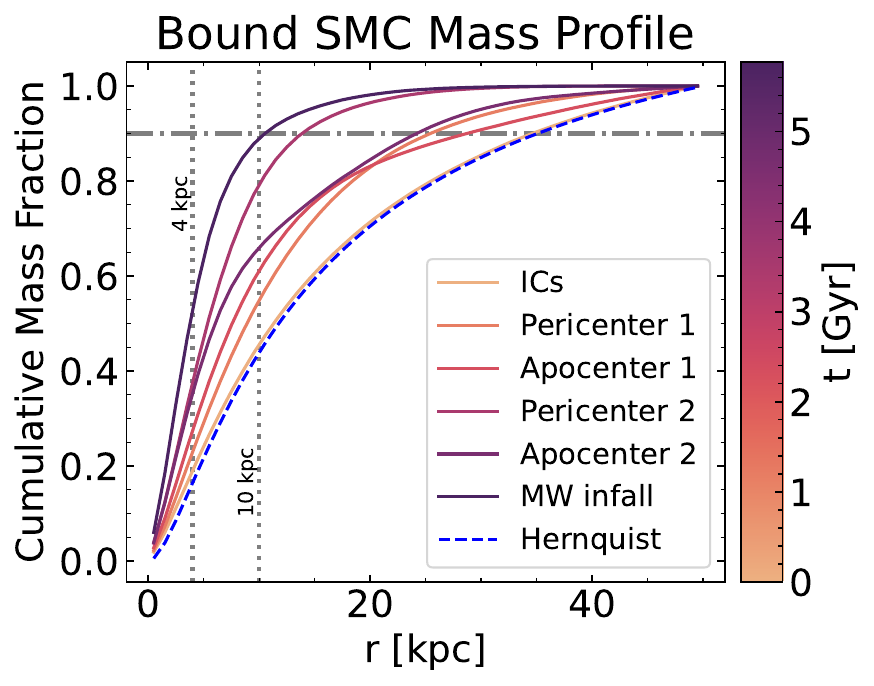}
    \caption{\textit{Left panel}: SMC bound mass enclosed within a 50 kpc radius from the SMC center as a function of time. Pericenter and apocenter times are marked with dotted lines. The SMC has lost roughly two-thirds of its initial DM mass to tidal stripping by the LMC -- retaining just $4.3 \times 10^9 \rm{M}_\odot$ at MW infall. \textit{Right panel}: evolution of the bound SMC's mass profile calculated from the simulation particles. Colored lines show the fraction of the SMC's bound mass enclosed within a radius $r$ (cumulative mass fraction), where each line corresponds to a time marked in the left panel. 90\% of the SMC's bound mass (dotted-dashed line) at MW infall is contained within a radius of $\approx10$ kpc.}
    \label{fig:smc_mass}
    
\end{figure*}

Panel \textit{d} of Figure \ref{fig:composite} shows the BFE-reconstructed density field of the SMC including its unbound DM debris. The SMC's debris at MW infall is composed of leading and trailing tidal arms embedded in a relaxed debris field that has been removed from the SMC during its previous pericenter passages. This extended debris field reaches $\sim100$ kpc from the LMC's center. The trailing arm overlaps with the SMC's wake, while the leading arm extends ahead of the SMC and overlaps with the Inner Overdensity.

The left panel of Figure \ref{fig:smc_mass} shows the SMC's mass loss as a function of time in the simulation, with pericenter times marked. By MW infall (the right edge of the plot), the LMC's tides have removed just over two-thirds of the SMC's original DM mass. The removed $\sim10^{10}\,\rm{M}_\odot$ of DM has joined the LMC halo and now contributes to the LMC's mass ($\sim6$\%). As such, the bound mass of the SMC at MW infall is just $4.3\times10^9\,\rm{M}_\odot$.

\subsection{The Bound SMC's Dark Matter Halo}\label{subsec:SMC_exp}

\begin{figure}
    \centering
    \includegraphics[width=\columnwidth]{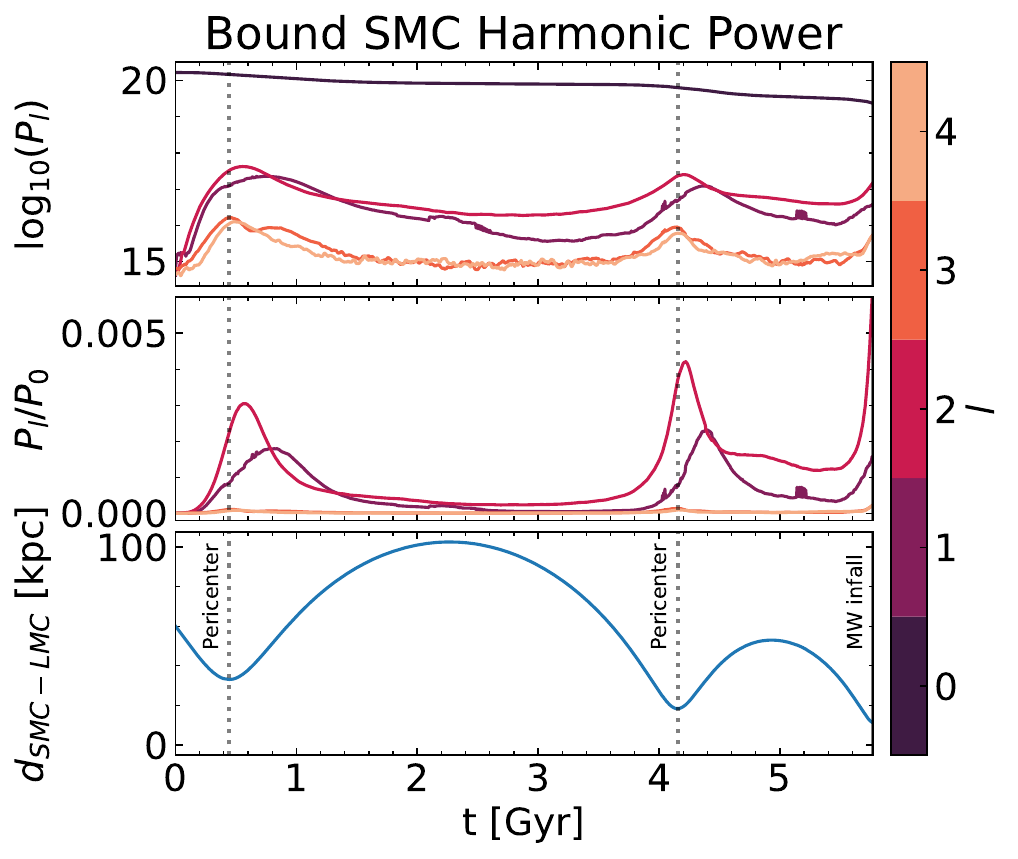}
    \caption{Time-evolution of the first five harmonic orders in the Bound SMC Expansion, following Figure \ref{fig:modes}. The Bound SMC Expansion shows dipole ($l=1$) peaks after SMC pericenters, although these peaks occur with roughly half the delay of the dipole peaks in the LMC Expansion. The noise in the SMC's dipole (e.g. at $t\approx5.2$ Gyr) is due to noise in the expansion centers; the \citet{power_inner_2003} centering algorithm (see Section \ref{subsec:sims}) is accurate to roughly twice the softening length ($\sim2\times80$ pc) when the SMC is highly distorted. The Bound SMC's quadrupole ($l=2$) shows peaks at pericenters that are an order of magnitude stronger than those in the LMC's quadrupole in Figure \ref{fig:modes}. The Bound SMC quadrupole corresponds to elongation due to the LMC's tides, whereas the LMC quadrupole captures the SMC's dynamical friction wake.  
    }
    \label{fig:modes_smc}
\end{figure}

\begin{figure}
    \centering
    \includegraphics[width=\columnwidth]{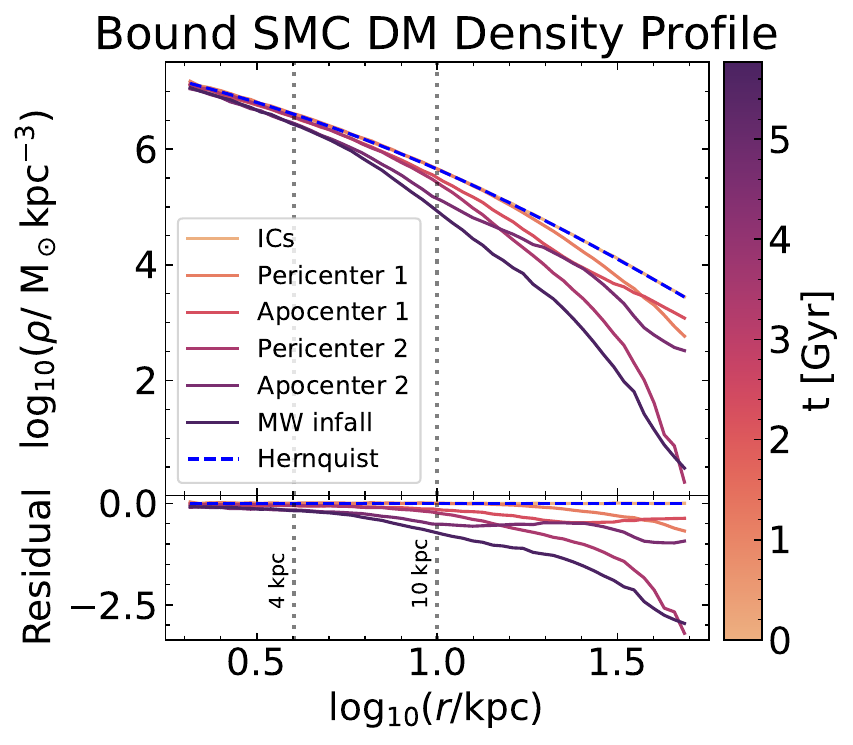}
    \caption{The Bound SMC's spherically averaged DM density profile, i.e. profiles are computed from particles bound to the SMC according to the definition in Section \ref{subsec:expansions}. The bottom panel illustrates the deviations from the IC, as in Figure \ref{fig:lmc_profiles}. The Bound SMC DM profile becomes steeper with time owing to mass loss from tidal stripping by the LMC. At MW infall, the slope of the density profile differs from the IC (Hernquist profile) by $> 10\%$ outside of 4 kpc. At 10 kpc, which encloses $\approx90\%$ of the SMC's bound mass (see Figure \ref{fig:smc_mass}), the slope of the Bound SMC's density profile is 66\% steeper than the Hernquist profile.}
    \label{fig:smc_profiles}
\end{figure}

\begin{figure}
    \centering
    \includegraphics[width=\columnwidth]{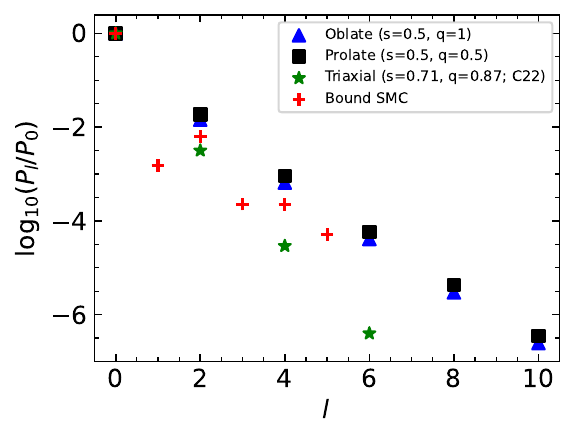}
    \caption{Gravitational power of each harmonic mode (normalized to the monopole) in the Bound SMC Expansion, compared to the oblate, prolate, and triaxial halos in Table \ref{tab:templates}. Tidal distortions from the LMC are evident as a peak in the quadrupole ($l=2$). The tidal distortions are not purely symmetric, as there is also power in the odd harmonics. In the even harmonics, the SMC has less power than the extreme oblate and prolate examples, but has more power in these terms than the cosmological triaxial halo of an LMC analog. The shape of the SMC's DM distribution at MW infall is not well described as spherical, oblate, prolate, or triaxial. 
    }
    \label{fig:template_power_smc}
\end{figure}

The density field of the Bound SMC Expansion at MW infall is shown in Panel (\textit{e}) of Figure \ref{fig:composite}. Like its debris, the bound SMC halo has been tidally stretched by the LMC, and is significantly elongated along roughly the SMC's direction of travel. 

To quantify the shape evolution of the SMC's halo, we plot the time-evolution of the power in the harmonics of the Bound SMC expansion in Figure \ref{fig:modes_smc}. Several features are notable, including (1) the decreasing strength of the monopole as the SMC loses mass; (2) the dipole peaks after pericenter passages with delays of 360 and 250 Myr for the first and second pericenter, respectively (roughly half the delay seen in the LMC); and (3) the quadrupole peaks at pericenter. The magnitude of the SMC's dipole peaks is comparable to the dipole peaks seen in the LMC relative to the monopole ($\sim0.002)$. However, the SMC's quadrupole peaks are nearly twice the relative amplitude of its dipole peaks (and an order of magnitude stronger than the LMC's quadrupole peaks). The quadrupole indicates significant tidal elongation of the Bound SMC by the LMC. At MW infall (the right edge of the plot), the SMC's quadrupole is at its maximum, indicating the highest level of tidal distortion seen during this simulation. As the SMC's orbit continues to decay to present-day, the quadrupole is expected to increase in strength.

Similar behavior has been seen in simulations of the LMC-MW system by \citet{lilleengen_effect_2023}, who reported that the LMC's quadrupole is excited by the MW's tides and is the strongest $l>0$ harmonic in the LMC during its recent pericenter with the MW. Again, this provides evidence for a consistent hierarchy of coefficient evolution during mergers with $\sim 1:10$ mass ratios, where (1) both the host and satellite experience dipole peaks after pericenter passages of the satellite due to their density center displacements, and (2) Both galaxies also experience quadrupole peaks at pericenter, although these have differing causes. The satellite's quadrupole dominates its expansion and is caused by the host's tides, while the host's quadrupole captures the satellite's dynamical friction wake and is weaker than its dipole. 

Figure \ref{fig:smc_profiles} shows the bound SMC's spherically averaged density profile at each pericenter, apocenter, and MW infall. LMC tides truncate the outskirts of the SMC's DM distribution over time. By MW infall (5.77 Gyr; black line), a Hernquist profile is a poor description of the SMC's overall density profile, as the true profile's slope differs from the original Hernquist profile by more than 10\% outside of $\sim 4$ kpc. The right panel of Figure \ref{fig:smc_mass} shows that 90\% of the SMC's bound mass at infall is contained within $\sim10$ kpc. This implies that the SMC's DM halo is in disequilibrium \textit{before} any interaction with the MW or its recent collision with LMC \citep[e.g.,][]{rathore_response_2025}, both of which are expected to perturb the SMC's halo even further \citep{rathore_galactic_2025}. 

Recently, \citet{rathore_response_2025} reported an estimate of the SMC's total mass within 2 kpc based on the torque it applies to the LMC's bar during the recent LMC-SMC collision. This estimate relied on the assumption that the SMC's halo is well described by a Hernquist sphere. Figure \ref{fig:smc_profiles} indicates that a Hernquist profile remains a reasonable description of the SMC's spherically averaged DM profile within 2 kpc at the time of MW infall, validating the calculation of \citet{rathore_response_2025}. Our result that the inner density profile of the SMC remains intact is consistent with observations of a DM cusp in the SMC by \citet{de_leo_surviving_2024}.

We compare the power in each Bound SMC expansion harmonic at MW infall to our template halos (see Table \ref{tab:templates}) in Figure \ref{fig:template_power_smc}. The Bound SMC Expansion has power in odd harmonics, indicating that its interaction with the LMC has produced asymmetries. However, the most striking feature of the Bound SMC's harmonic amplitudes is the strength of the quadrupole caused by the LMC's tides, which reaches just under 1\% of the strength of the monopole. Overall, the strength of the Bound SMC's harmonics lies between the cosmological expectation (with the caveat that this is the expectation for an LMC-mass halo) and the extreme prolate and oblate examples. 

The shape of the Bound SMC’s DM distribution at MW infall is not well described as spherical, oblate, prolate, or triaxial. In particular, the LMC's tides have perturbed the SMC's shape beyond that typically expected from a cosmological assembly history. 

\subsection{The Combined LMC-SMC Dark Matter Distribution}\label{subsec:comb_exp}

\begin{figure*}
    \centering
    \includegraphics[width=\textwidth]{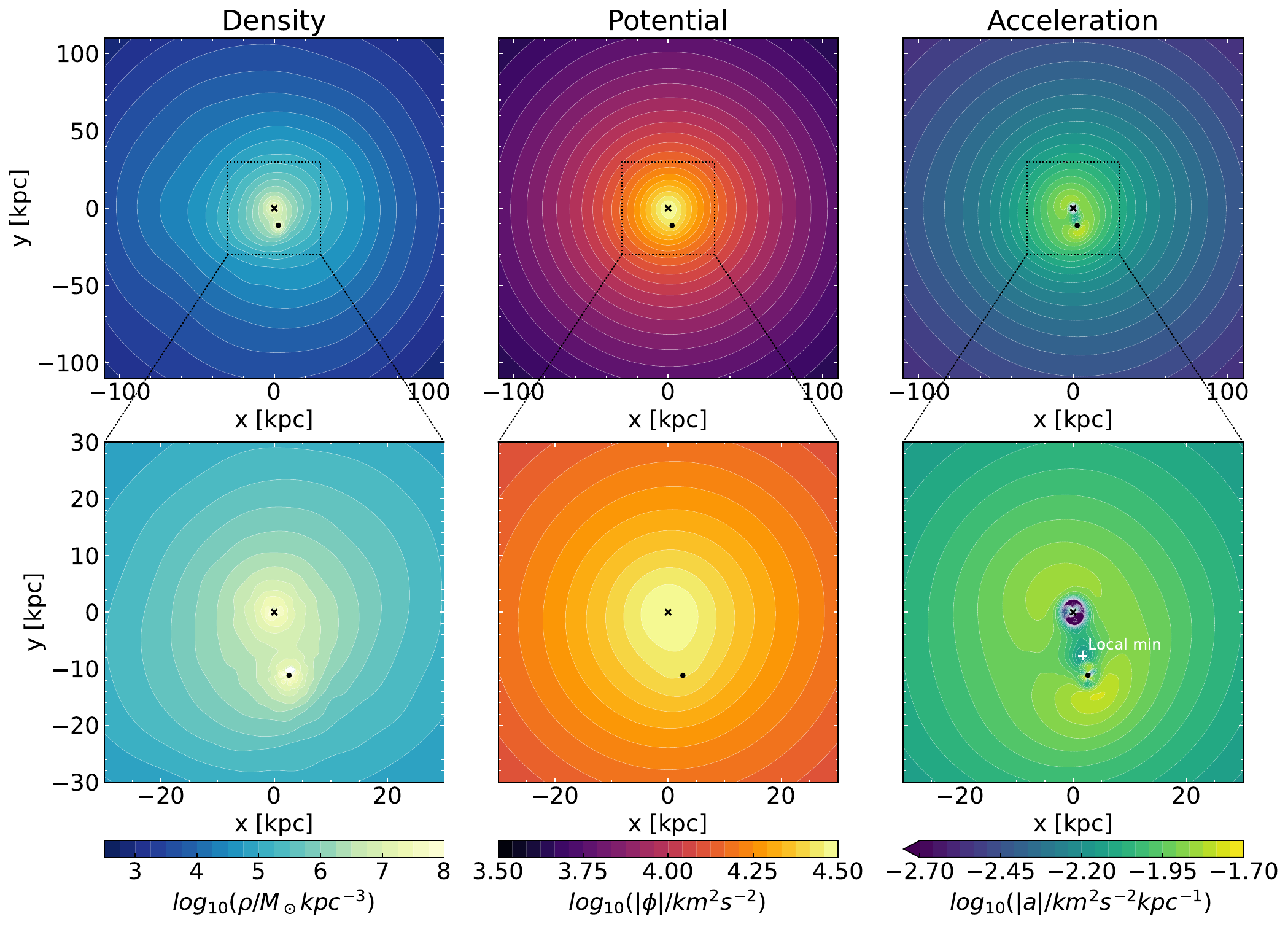}
    \caption{Density (\textit{left column}), potential (\textit{center column}), and acceleration (\textit{right column}) fields of the combined LMC+SMC Expansions at MW infall (i.e. the top left panel is the same as panel (\textit{a}) of Figure \ref{fig:composite}). Panels within a column share a color scale. The bottom row of panels show zoom-ins of the inner 30 kpc of the top panels. All fields show asymmetries caused by the LMC and SMC's mutual distortions, which are most apparent in the zoomed-in bottom row of panels. There is a local minimum in the acceleration field (marked with a cross) between the Clouds, approximately 7.9 (3.6) kpc away from the LMC (SMC). This minimum corresponds to the $L_1$ Lagrange point between the Clouds, and determines the SMC's tidal radius. 
    }
    \label{fig:fields}
\end{figure*}

In this section, we discuss the DM distribution of the LMC-SMC system by combining the LMC Expansion with the SMC Expansion. Figure \ref{fig:fields} shows the DM density (same as Panel (\textit{a}) of Figure \ref{fig:composite}), potential, and magnitude of the acceleration field of the LMC-SMC system at MW infall.

Within $\sim$20 kpc of the LMC center, the DM potential and density are dominated by the LMC and SMC's cusps, which elongate the isopotential and isodensity contours in the direction of the SMC. A notable feature of the acceleration field is the $L_1$ Lagrange point (local acceleration minimum) between the Clouds, approximately 7.9 (3.6) kpc from the LMC (SMC). This Lagrange point sets the tidal radius of the SMC, and is consistent with the radius at which the SMC's density profile deviates from Hernquist in Figure \ref{fig:smc_profiles}. The acceleration field is also asymmetric around this minimum owing to the elongated Bound SMC halo (panel (\textit{e}) of Figure \ref{fig:composite}) and the shape of the SMC's tidal arms (panel (\textit{d}) of Figure \ref{fig:composite}). 

At larger distances (upper panels), deviations from sphericity are driven by the SMC's debris field (which elongates the isodensity contours along the $-x$-axis) and perturbations to the LMC halo from the SMC.

\subsubsection{Consequences for Orbit Integration within the LMC-SMC System}\label{ssubsec:force_errors}

\begin{figure}
    \centering
    \includegraphics[width=\columnwidth]{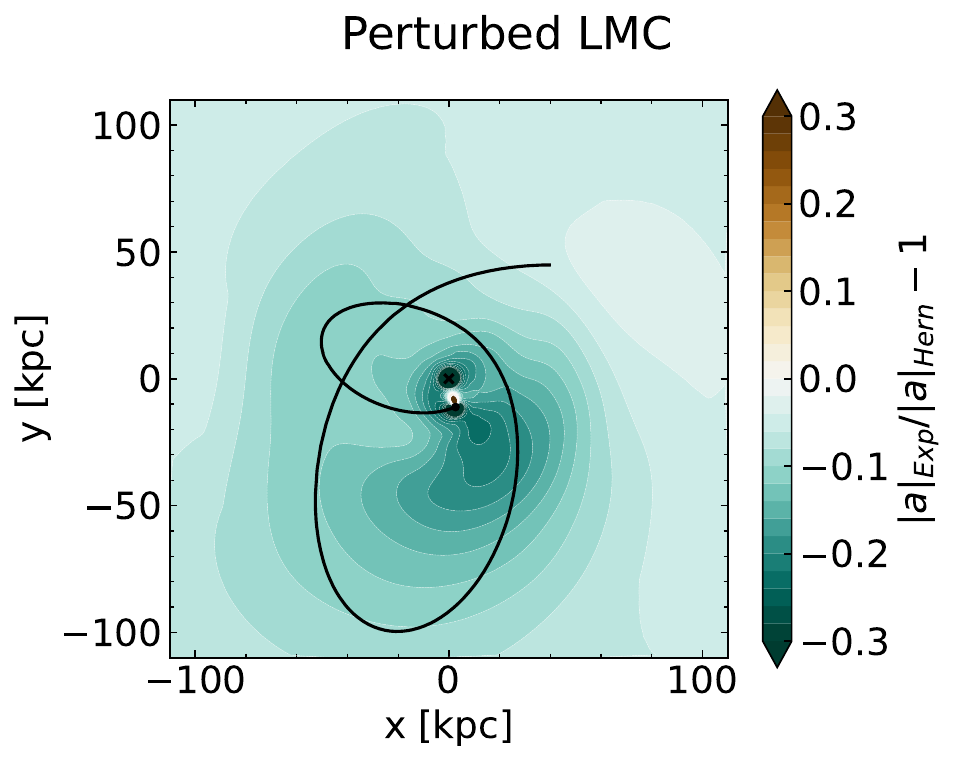}
    \includegraphics[width=\columnwidth]{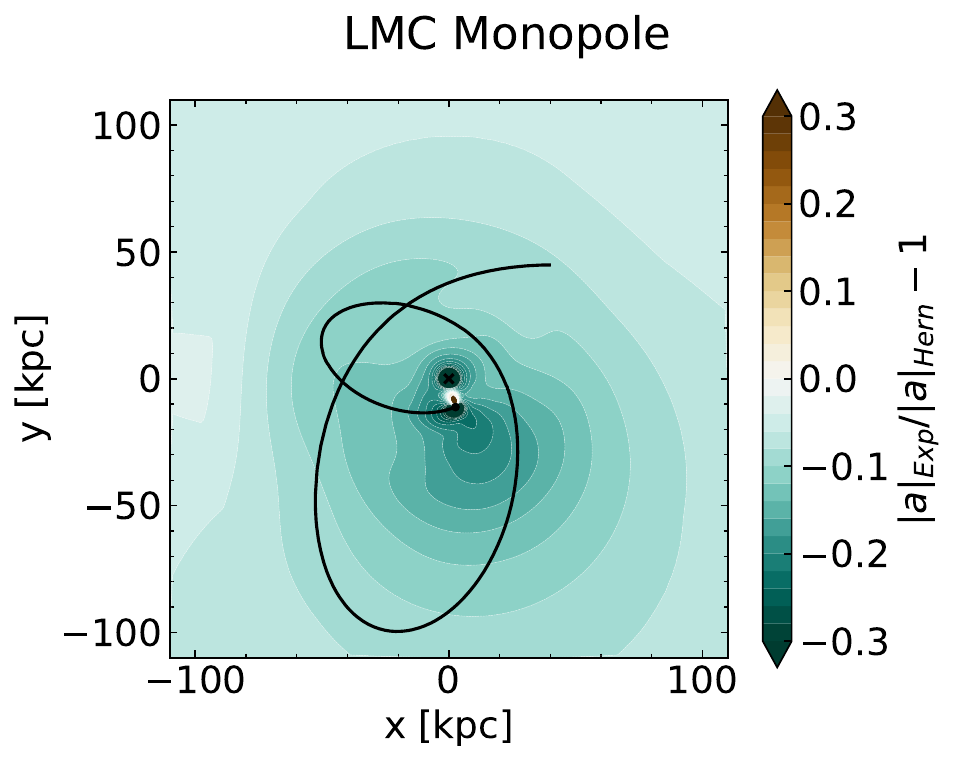}
    \caption{Relative error in the acceleration field of the LMC-SMC system at MW infall calculated in the SMC orbital plane with BFEs vs. the spherical potentials used to create the LMC and SMC ICs (see Equation \ref{eqn:force_errors}). The SMC's orbit from our \textit{N}-body simulation is included as a reference. \textit{Top panel:} the BFE accelerations are calculated from the combination of the LMC Expansion and SMC Expansion (panel (\textit{a}) of Figure \ref{fig:composite}). The rigid-sphere approximation: (1) overestimates the strength of the BFE acceleration field by up to 20\% (e.g. 15 kpc $\lesssim r \lesssim$ 80 kpc at negative $y$); and (2) incorrectly identifies the local minimum between the LMC and SMC by approximately 3 kpc. \textit{Bottom panel:} the BFE accelerations are calculated from the LMC's $l=0$ terms plus the full SMC Expansion, isolating the impact of the SMC's shape/debris on the acceleration field. While the magnitudes of the force errors are comparable to those using the full LMC Expansion, the shape of the force errors changes, demonstrating that \textit{both} the LMC and SMC's shapes must be accounted for when computing orbits of halo tracers in this system.}
    \label{fig:force-errors}
\end{figure}

The accuracy of the adopted acceleration field has consequences for modeling several observed features of the LMC-SMC system via orbit integration. In particular, there is evidence that the LMC possesses an extended stellar halo \citep{minniti_kinematic_2003, alves_stellar_2004, majewski_discovery_2009, saha_first_2010, belokurov_clouds_2017, nidever_exploring_2019, petersen_tidally_2022} with rich substructure including stellar streams \citep{belokurov_stellar_2016, navarrete_stellar_2019, zaritsky_untangling_2025} and tidal features from its interaction with the SMC \citep[e.g.,][]{balbinot_lmc_2015, besla_low_2016, mackey_10_2016, belokurov_clouds_2017, cullinane_magellanic_2020, chandra_discovery_2023, munoz_chemo-dynamical_2023}. Additionally, the LMC likely fell into the MW with its own population of satellites, in addition to the SMC \citep{donghia_small_2008, jethwa_magellanic_2016, sales_identifying_2017, kallivayalil_missing_2018, erkal_limit_2020, patel_orbital_2020, battaglia_stellar_2022, correa_magnus_measuring_2022, pace_proper_2022}. Finally, several hypervelocity stars have been identified as likely originating from the LMC \citep{erkal_hypervelocity_2019, han_hypervelocity_2025, lucchini_threading_2025}.

When building models to explain these observations, the common approach is to treat the LMC and SMC DM halos as spherically symmetric and static. That is, while the halos of the Clouds are allowed to move in response to each other's gravity, they do not change shape or lose mass over time. This ``rigid-potential'' approach is computationally efficient and effective at capturing the effects of a noninertial reference frame on tracer orbits \citep[e.g.,][]{gomez_and_2015}. However, in contrast to \textit{N}-body or BFE methods, rigid-potentials cannot capture asymmetric shape distortions of the halos that can affect the acceleration field. 

We quantify the importance of DM halo asymmetries to the forces experienced by tracers orbiting in the LMC-SMC system in Figure \ref{fig:force-errors}. The top panel shows the relative error between the LMC+SMC Expansion-reconstructed acceleration field ($|a|_{Exp}$) and the acceleration field of the two Hernquist spheres used to generate the ICs ($|a|_{Hern}$), given by

\begin{equation}\label{eqn:force_errors}
    \delta a = \frac{|a|_{Exp}}{|a|_{Hern}}-1
\end{equation}

\noindent We note that the basis does not resolve the inner $\sim 1$ kpc of either halo, so deviations between the two fields near the halo centers are an artifact. However, outside of these regions, discrepancies between the two acceleration fields are still apparent. In particular, the rigid-potentials fail to account for the distortions of both halos, incorrectly locate the local minimum between the LMC and SMC cusps (SMC tidal radius) by $\sim 3$ kpc, and overestimate the acceleration field by up to 20\%, depending on the location. 

Our results in Section \ref{subsec:LMC_exp} show that the spherically averaged density profile of the LMC does not evolve throughout the simulation. This implies that it may be appropriate to ignore the SMC's dynamical friction wake and other perturbative effects. To test whether treating the LMC as spherical is a reasonable approximation, we plot the error in the acceleration field with respect to the rigid-potentials using the LMC Expansion's monopole ($l=0$) terms plus the full SMC Expansion (bottom panel of Figure \ref{fig:force-errors}). We find that the shape of the force error field changes significantly when only using the spherically symmetric terms of the LMC Expansion, demonstrating that the SMC-driven perturbations in the LMC's halo also affect the acceleration field.

We conclude that the distortions in both the SMC and LMC halos cannot be ignored when computing the trajectories of objects in orbit within the combined system. 

\subsubsection{Consequences for Modeling the SMC's Orbit about the LMC}

\begin{figure*}
    \centering
    \includegraphics[height=3.1in]{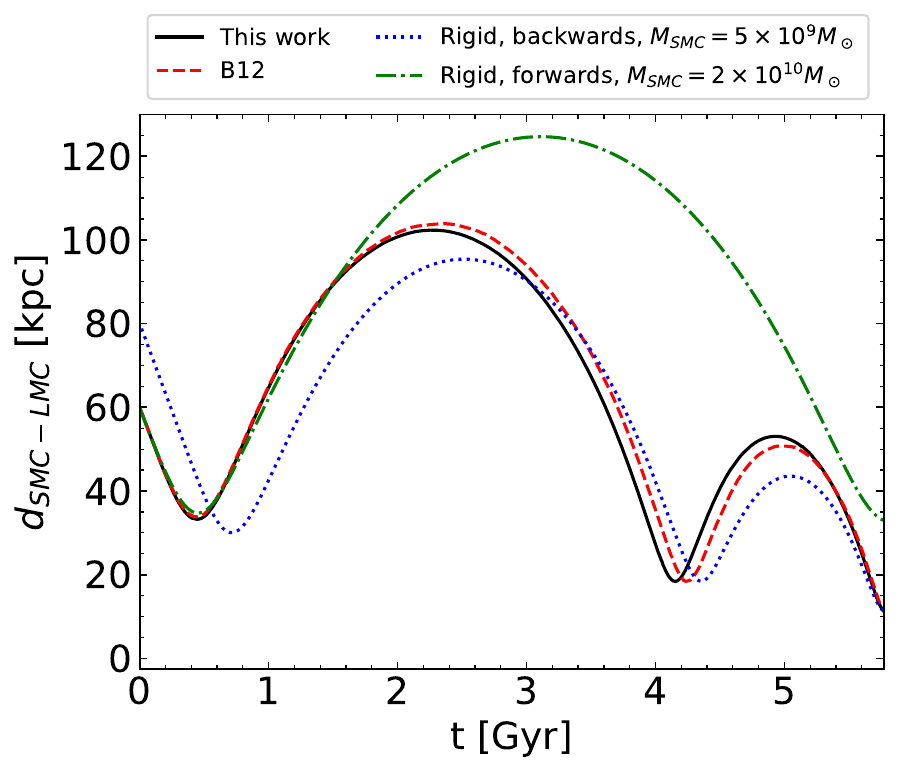}
    \includegraphics[height=3.1in]{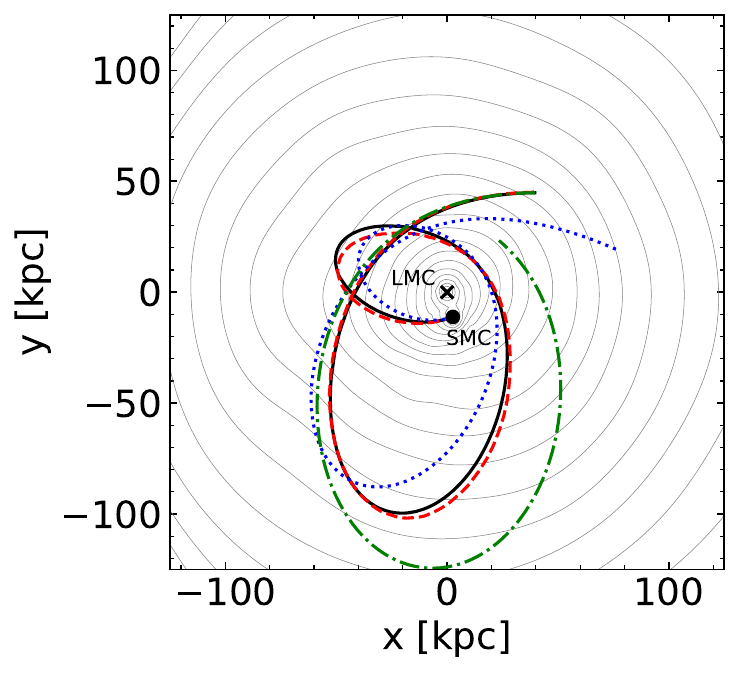}
    \caption{\textit{N}-body LMC-SMC orbits compared to integrations with analytic potentials. \textit{Left panel}: same as Figure \ref{fig:orbit_comp} with the addition of orbits integrated using rigid analytic potentials (see text for details). We use the method of \citet{patel_orbital_2020} for the LMC-MW interaction to integrate this orbit forwards (dot-dashed lines) or backwards (dotted lines) in time, including the drag force from dynamical friction. \textit{Right panel}: LMC-SMC \textit{N}-body and rigid-potential orbits projected in the SMC orbital plane. 
    Contours show the DM density field of the combined system at MW infall (as in panel (\textit{a}) of Figure \ref{fig:composite}). Both orbits obtained by integrating analytic potentials are a poor match to the orbits obtained from \textit{N}-body simulations.}
    \label{fig:analytic_orbits}
\end{figure*}

Orbits obtained by integrating rigid-potentials often diverge from orbits produced in \textit{N}-body simulations on timescales longer than an orbital period \citep[e.g.,][]{van_der_marel_m31_2012-1}. To understand this effect in the LMC-SMC system, in Figure \ref{fig:analytic_orbits} we compare the LMC-SMC orbit in two \textit{N}-body simulations (this work and \Bref{}) to orbits obtained by integrating rigid-potential models of the LMC and SMC using the methods of \citet{patel_orbital_2020} for the LMC-MW system. 

The rigid LMC model includes a \citet{miyamoto_three-dimensional_1975} stellar disk and a Hernquist DM halo. The disk potential is designed to closely match the rotation curve of the LMC disk in our \textit{N}-body simulation, with a total mass of $3.6\times10^9\,\rm{M}_\odot$, and a scale length (height) of 1.7 (0.34) kpc. The LMC DM halo parameters match those used to generate the IC of our \textit{N}-body LMC simulation as outlined in Section \ref{subsec:sims}.

Meanwhile, we model the SMC as only a Hernquist DM halo, and investigate two choices for its mass: 

\begin{enumerate}
    
    \item Our first SMC DM halo model is designed to match our \textit{N}-body simulation IC. The analytic SMC halo matches the Hernquist profile used to generate the SMC IC as described in Section \ref{subsec:sims}, i.e. the SMC's mass is $2\times10^{10}\, \rm{M}_\odot$. The orbit of this SMC model is integrated forward in time for 5.77 Gyr, starting from the same phase-space coordinates as our \textit{N}-body simulation (see Table~\ref{tab:init_phase}).

    \item Our second SMC model is designed to represent the SMC at MW infall, i.e. the SMC's mass is $5\times10^9\, \rm{M}_\odot$. The orbit of this SMC model is integrated backwards in time for 5.77 Gyr from the SMC's phase-space position at MW infall. 
\end{enumerate}
 
In our rigid-potential orbits, we include the effects of dynamical friction on the SMC. \citet{van_der_marel_m31_2012-1} attempted to reconcile rigid-potential integrations of M33's orbit about M31 with \textit{N}-body simulations by adopting a tuned parameterization of the Coulomb logarithm. This scheme was later adopted by \citet{patel_orbital_2020} to integrate the LMC's orbit about the MW. As both of the M31-M33 and MW-LMC systems have a mass ratio of $\sim1:10$ like our LMC-SMC system, we adopt the same prescription for dynamical friction in our rigid-potential integrations.\footnote{Note that \citet{patel_orbital_2020} used a different dynamical friction formalism and Coulomb logarithm for their LMC-SMC orbit.}

Figure \ref{fig:analytic_orbits} shows that, even when using a dynamical friction scheme tuned to 1:10 mass-ratio mergers, LMC-SMC orbits obtained by integrating rigid-potentials are a poor match to \textit{N}-body simulations. This can be explained by two primary factors. First, the rigid-potentials cannot account for the significant mass loss of the SMC during the \textit{N}-body simulation (Figure \ref{fig:smc_mass}). Second, the rigid-potentials do not capture the LMC's density center displacement, which depends strongly on time and radius from the LMC center (Figure \ref{fig:evolution}). 

Caution is required when trying to recover the SMC's orbital history with rigid-potentials. The orbits presented here are a best-case scenario, where we have used a dynamical friction scheme tuned specifically for our LMC:SMC mass ratio. These errors will only be exacerbated if a different mass ratio is assumed. Additionally, at MW infall, the SMC is still well separated from the LMC and the perturbations in the LMC halo are relatively weak (see Figure \ref{fig:modes}). As the merger progresses toward the present-day and the SMC collides with LMC, the rigid-potential description of the Clouds will become an even poorer approximation.

Ultimately, our results imply that using a rigid-potential model for either or both galaxies can result in significantly different orbits for objects in the LMC-SMC system and the Clouds themselves compared to \textit{N}-body or BFE-based methods. This has implications for simulations of the Stream's formation, LMC halo stars, stellar streams around the LMC, pre-MW infall orbits of LMC satellites, and hypervelocity stars originating from the LMC, suggesting that simulation frameworks that account for halo distortions and self-gravity are needed to describe these features accurately. 

\section{Discussion}\label{sec:disc}

In this section, we discuss the implications of our results in several astrophysical contexts. These include the effect of the SMC on the LMC's orbit about the MW (Section \ref{subsec:smc_wake}), predictions for the LMC's reflex motion (Section \ref{subsec:reflex}), the effect of the LMC's halo distortions on the LMC disk (Section \ref{subsec:lmc_disk}), the nature of DM (Section \ref{subsec:dm_physics}), and interpreting observations of isolated dwarf galaxy pairs (Section \ref{subsec:dwarf_pairs}).

\subsection{The SMC's Effect on the LMC's Wake in the MW}\label{subsec:smc_wake}

The LMC's dynamical friction wake in the MW is a key observable prediction of the LMC-MW interaction \citep[e.g.,][]{garavito-camargo_hunting_2019, petersen_reflex_2020} that can be used to constrain the LMC's mass and orbit \citep[e.g.,][]{conroy_all-sky_2021, amarante_mapping_2024, fushimi_determination_2024, cavieres_distant_2025}, and used as a probe of DM particle physics \citep{foote_structure_2023}. Typically, models of the LMC's wake have not explicitly accounted for the presence of the SMC. Instead, some authors choose a range of LMC masses, arguing that the SMC's influence on the wake is effectively captured by raising the mass of the LMC \citep[e.g.,][]{garavito-camargo_hunting_2019}. 

Our results show that the SMC is significantly tidally stripped by the LMC by the time the system falls into the MW halo. In our simulations, the SMC's DM mass is $4.3\times10^9\,\rm{M}_\odot$ at MW infall, just over 2\% of the LMC's DM mass. Based on these findings, we argue that the approach of accounting for the SMC's mass by varying the LMC mass in models of the LMC's wake formation is justified. 

\subsection{Reflex Motion of the LMC}\label{subsec:reflex}

\begin{figure}
    \centering
    \includegraphics[width=\columnwidth]{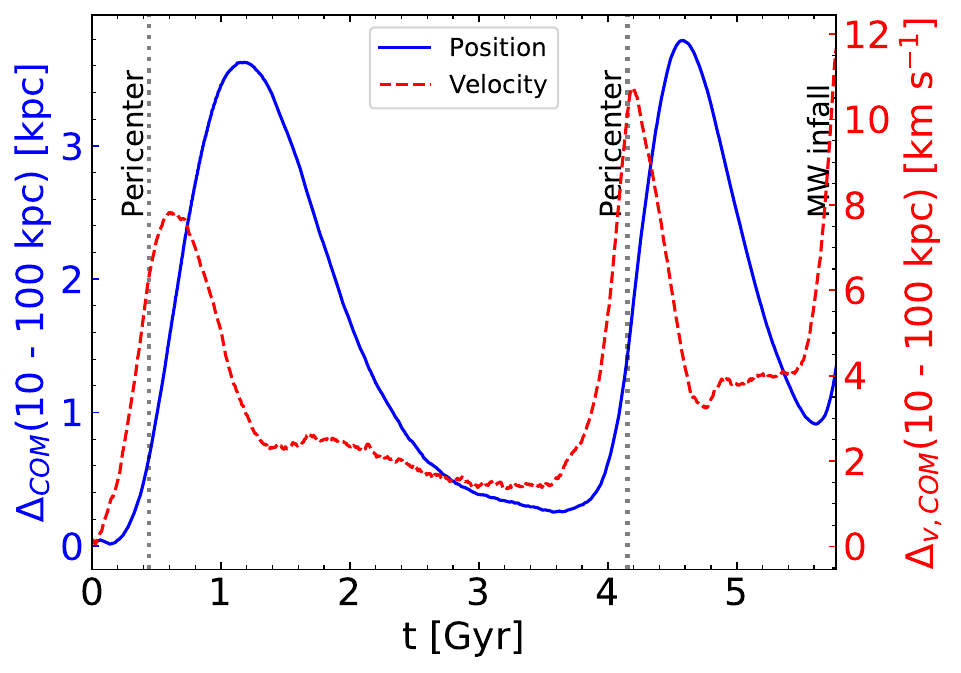}
    \caption{Center of mass (COM) displacement of the LMC DM halo in position ($\Delta_\mathrm{COM}$; solid line) and velocity (($\Delta_{v,\mathrm{COM}}$; dashed line) as a function of time. Both quantities are calculated by finding the difference in the COM position/velocity when measured using the LMC's inner halo ($< 10$ kpc) vs. outer halo ($<100$ kpc). The LMC's inner halo is displaced from the outer halo by approximately 3.3-3.4 kpc following each pericenter passage of the SMC. The density center displacement is tightly correlated to the strength of the dipole power in the LMC expansion (compare to the $l=1$ line in the middle panel of Figure \ref{fig:modes}). The density center displacement will result in an apparent reflex motion of the outer halo relative to an observer in the LMC disk (see Figure \ref{fig:rv_shells}). This effect is qualitatively similar to the density center displacement of the MW outer halo due to the recent pericentric approach of the LMC. The velocity of the inner halo with respect to the outer halo reaches peaks of $\approx10$ km s$^{-1}$ during SMC pericenter passages. }
    \label{fig:com_disp} 
\end{figure}

\begin{figure*}
    \centering
    \includegraphics[width=\textwidth]{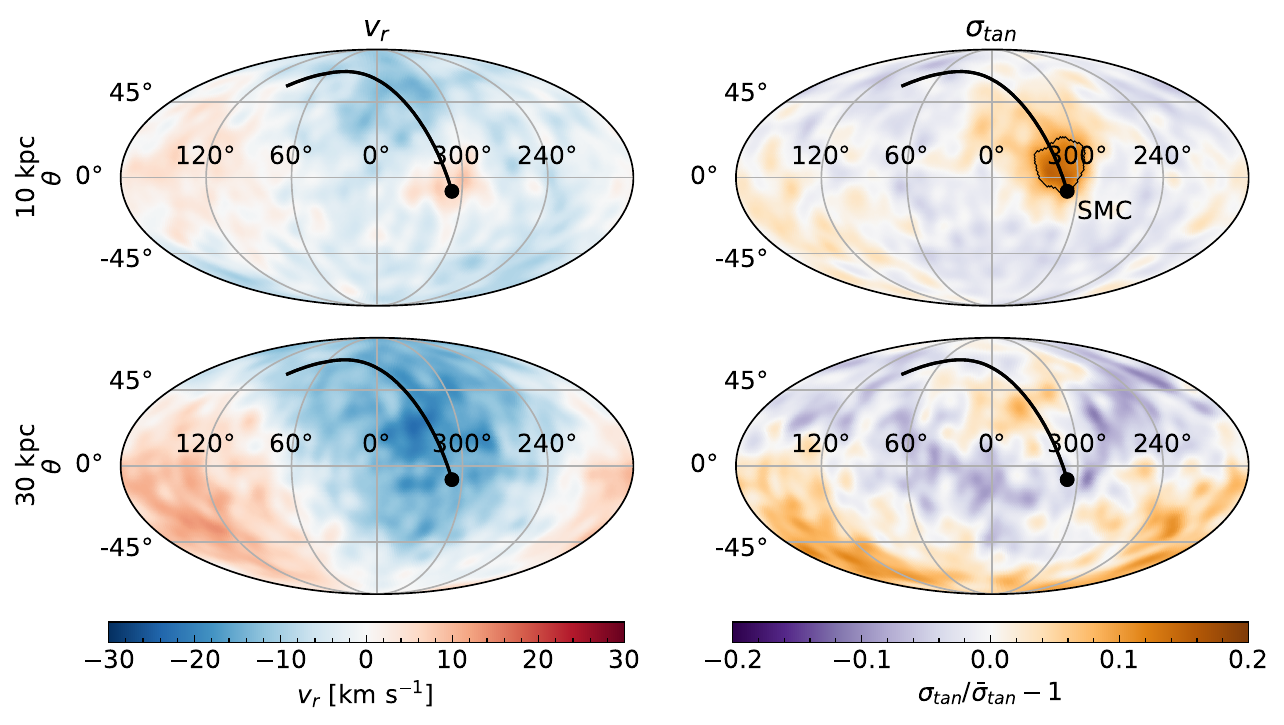}
    \caption{Mollweide projections in the LMC disk plane, illustrating the mean radial velocity (\textit{left panels}) and tangential velocity dispersion contrast ($\sigma_\mathrm{tan}/\bar{\sigma}_\mathrm{tan}-1$; \textit{right panels}) of LMC DM halo particles at the time of MW infall. Each row shows particles in 5 kpc thick spherical shells centered at 10 kpc (top) and 30 (bottom) kpc from the LMC center. All projections are in the LMC's disk plane, i.e. $\theta=0$ is aligned with the LMC's disk, and all velocities are with respect to the LMC's center. The dot shows the SMC's projected location at MW infall ($\sim$12 kpc from the LMC). The solid line shows the SMC's projected orbit since its last apocenter 831 Myr ago. \textit{Left column:} there is a clear radial velocity dipole of $\sim 15$ km s$^{-1}$ in the LMC halo at distances larger than the SMC's location ($>$10 kpc). This is similar to the radial velocity dipole (i.e. reflex motion) observed in the MW due to the LMC \citep[e.g.,][]{yaaqib_radial_2024}. \textit{Right column:} at 10 kpc, there is a $\sim 15 \%$ spike in the tangential velocity dispersion trailing the SMC, which is a signature of the SMC's wake (see \citet{garavito-camargo_hunting_2019} for the corresponding signature in the MW due to the LMC). A contour is placed at a dispersion contrast of 0.1 in the upper right panel, which traces the wake boundary. While our goal is not to make direct observational predictions, our results indicate that the velocity signatures we identify in the LMC halo (radial velocity dipole and wake) are generic in 1:10 mergers and result primarily from the most recent pericenter passage of the satellite.}
    \label{fig:rv_shells}
\end{figure*}

The clearest observed signal of the MW's density center displacement caused by the LMC is a dipole in the radial velocities of MW halo stars. The radial velocity dipole is the reflex motion of the inner MW halo about the LMC-MW barycenter \citep{erkal_detection_2021, petersen_detection_2021, yaaqib_radial_2024, brooks_milky_2025, brooks_simulation-based_2026, bystrom_exploring_2025, chandra_all-sky_2025}. The LMC is not the only large satellite currently merging with the MW, however -- Sgr has completed multiple orbits around the MW over $\sim 6$ Gyr \citep[e.g.,][]{laporte_influence_2018}. Sgr should also leave velocity signatures in the MW halo, analogous to those left by the LMC. However, Sgr's orbital history differs significantly from the LMC's, where the latter is on first infall. It is unclear how velocity signatures in the MW stellar and DM halo left by Sgr may have evolved over its multiple pericenter passages. These velocity signatures include dipoles in radial velocity and increase in the tangential velocity dispersion in the vicinity of DM wakes \citep{garavito-camargo_hunting_2019}. 

\citet{vasiliev_dear_2024} studied radial velocity dipole signatures in the MW DM halo in the context of a second-infall orbit for the LMC, finding that the MW reflex motion in this scenario was nearly identical to a first-infall model for the LMC orbit. 

In this paper, we have shown that density perturbations induced in the LMC's DM halo over multiple SMC pericenter passages are more complex than first-infall orbits, owing to successive dipole excitations during each pericenter passage. Here, we explore velocity signatures in the LMC DM halo shaped by multiple SMC pericenter passages, extending the work of \citet{vasiliev_dear_2024} to a lower host-mass scale and more frequent pericenter passages of the satellite. We emphasize that our goal is \textit{not} to predict present-day observables. Instead, we seek a qualitative comparison between the velocity signatures in our simulation and those identified in MW-LMC simulations \citep[e.g.,][]{garavito-camargo_hunting_2019, vasiliev_dear_2024}.

We begin by quantifying the density center displacement of the LMC DM halo over time in Figure \ref{fig:com_disp}. The displacement of the LMC's inner DM halo with respect to the outer DM halo is calculated as the displacement between the LMC's center of mass (COM) when measured with DM particles inside 10 vs. 100 kpc of the LMC's expansion center. The inner and outer DM halos are displaced from each other in position (left axis; solid line) by up to 3.4 kpc, with two peaks in the displacement following the SMC's pericenter passages. These peaks have approximately the same magnitude at both pericenter passages, despite the SMC's mass loss between its first and second pericenter. This can be explained when considering the second pericenter is lower ($\approx18.5$ kpc) than the first pericenter ($\approx33.2$ kpc) -- the ratio of the SMC's mass to the LMC's mass enclosed by the pericenter distance is roughly 1:5 at both pericenter times. In general, the behavior of the density center displacement is tightly correlated to the strength of the LMC's dipole in Figure \ref{fig:modes}, consistent with findings in cosmological simulations by \citet{darragh-ford_shaping_2025}. 

We next search for velocity signatures in the LMC's DM halo particles at MW infall, taking predicted signatures seen in MW-LMC simulations by \citet{garavito-camargo_hunting_2019} as a guide. Velocities are computed in an LMC-centered frame, i.e. as seen by an observer at the center of the LMC. In the left column of Figure \ref{fig:rv_shells}, we plot the radial velocity of DM particles in 5-kpc thick shells at two distances, 10 and 30 kpc. The inner shell is just inside the SMC's 12 kpc separation from the LMC at the time of MW infall. 

At 10 kpc, the radial velocity field is relatively unperturbed. A clear velocity dipole of $\sim 15$ km s$^{-1}$ is seen in the 30 kpc shell, with positive velocities leading the SMC and negative velocities trailing. In Figure \ref{fig:com_disp}, we also plot the difference between the LMC's COM velocity when measured with particles inside 10 vs. 100 kpc (right axis; dashed line). The velocity displacement peaks near SMC pericenter passages. As such, the LMC's velocity dipole is likely near a maximum at the epoch shown in Figure \ref{fig:rv_shells}.

The exact geometry and orientation of the dipole will depend on the orientation of the SMC's orbit with respect to the LMC's disk, as the dipole is excited in the SMC's orbital plane. Here, the SMC's orbit is roughly polar with respect to the LMC's disk, and we expect that alternative orbital scenarios for the SMC will yield different orientations for this dipole. Further details on the SMC's orbit geometry in our simulation will be released in our companion paper (H. Rathore et al. in preparation).

The right column of Figure \ref{fig:rv_shells} shows the tangential velocity dispersion (relative to the mean of the shell) of DM halo particles in the same fashion as the left panel. Here, the primary signature is an increase in the dispersion in the SMC's wake, which is most apparent in the 10 kpc shell.

We caution against extrapolation of these results to direct present-day observables. In particular, we identify the reflex dipole and SMC wake at MW infall, and the subsequent evolution of the combined MW-LMC-SMC system will result in different signatures at present day. Additionally, our analysis relies only on LMC DM particles (as opposed to a realistically constructed stellar halo), without ``tagging'' DM particles to match a stellar density profile as was done by, e.g. \citet{garavito-camargo_hunting_2019}. A more realistic model for the LMC's stellar halo that includes the correct density profile and substructure may complicate these signatures \citep{cunningham_quantifying_2020}.

Nevertheless, both of the velocity signatures we identify in the LMC halo due to the SMC are directly analogous to those predicted in the MW due to the LMC on both first-infall \citep{erkal_total_2019, garavito-camargo_hunting_2019, cunningham_quantifying_2020, petersen_reflex_2020, vasiliev_effect_2023, vasiliev_dear_2024, sheng_uncovering_2024, sheng_lmc-induced_2025} and second-passage orbits \citep{vasiliev_dear_2024}, which have been observed by several authors \citep{erkal_detection_2021, petersen_detection_2021,  yaaqib_radial_2024, brooks_milky_2025, brooks_simulation-based_2026, bystrom_exploring_2025, chandra_all-sky_2025}. Our results demonstrate that radial velocity dipoles in a host halo, due to the displacement of its density center by a satellite, can be seen in $\sim1:10$ mergers with both MW- and LMC-mass hosts.

Furthermore, we find that velocity perturbations in the halo primarily encode information from only the most recent pericenter passage of the satellite. As such, we may expect that any velocity signatures left by Sgr in the MW reflect only its recent orbital history. 

\subsection{Implications for the LMC's Disk}\label{subsec:lmc_disk}

The LMC's disk is highly perturbed, containing a single spiral arm \citep[e.g.,][]{de_vaucouleurs_structure_1972, besla_low_2016, mackey_10_2016}, an offset and tilted stellar bar \citep[e.g.,][]{choi_smashing_2018, jimenez-arranz_vertical_2025, rathore_precise_2025}, and warped outskirts \citep[e.g.,][]{van_der_marel_magellanic_2001, choi_smashing_2018, saroon_shape_2022, jimenez-arranz_vertical_2025}. Some of these features, such as the unusual position and low pattern speed of the bar, can be traced directly to torques from the SMC (e.g., \Bref{}; \citealt{jimenez-arranz_kratos_2024, jimenez-arranz_tidal_2025, rathore_response_2025}). 

We have shown that the LMC's own DM halo has been significantly perturbed by the SMC, opening the possibility that the asymmetries in the LMC's halo may also be partially responsible for the curious structure of the LMC's visible components. Indeed, \citet{weinberg_production_1995, weinberg_dynamics_1998} have demonstrated that satellite-induced resonant modes in a DM halo can induce warps in a host galaxy's disk, although this process is sensitive to the details of the satellite (see also \citealt{gomez_fully_2016} for a demonstration of this effect in a cosmological simulation). Quantifying the contributions to the LMC disk's morphology of halo distortions induced by the SMC vs. the SMC's direct torques will provide some of the most detailed evidence yet of DM's role in shaping merging galaxies. A detailed exploration of this question with the simulation used in this paper is the subject of ongoing work to be released in a companion paper (H. Rathore et al. in preparation). 

\subsection{Implications for Dark Matter Particle Physics}\label{subsec:dm_physics}

In Section \ref{subsec:SMC_exp}, we showed that the SMC's halo structure varies in a time-dependent manner due to the LMC's tides in a cold DM simulation, losing mass to tidal stripping and developing a quadrupole whose strength varies along the SMC's orbit. This scenario offers a unique opportunity to constrain DM particle physics in the LMC-SMC system, as tidal stripping of DM halos is sensitive to the DM particle model assumed. 

For instance, ultralight or fuzzy DM solitons can undergo runaway tidal disruption \citep{du_tidal_2018}, depending on the properties of the halo's outskirts \citep{schive_soliton_2020}. Introducing a self-interaction term to an ultralight DM model can also affect stripping times \citep{glennon_tidal_2022, dave_uldm_2024}. In addition, \citet{berezhiani_thermalization_2023} found that tides from a host can inhibit soliton formation in superfluid DM halos. In self-interacting DM models with particle masses comparable to cold DM, cored density profiles can make satellites more susceptible to disruption \citep[e.g.,][]{nadler_signatures_2020, nadler_sidm_2023, correa_constraining_2021}. Tides can also accelerate the core-collapse process of a self-interacting DM halo \citep[e.g.,][]{nishikawa_accelerated_2020, yang_self-interacting_2020, zeng_core-collapse_2022}. Clearly, there are many interesting physical effects on the tidal evolution of a subhalo that depend on the assumed nature of the DM particle. 

Improving the observational understanding of the SMC's DM mass profile is key to unlocking the SMC's potential to discriminate between these DM models. Recent works have made advances toward constraining the SMC's inner mass profile while accounting for disequilibrium. For example, \citet{de_leo_surviving_2024} carefully applied Jeans modeling to SMC stars identified in \textit{Gaia} DR3 while accounting for and removing unbound stars, obtaining a constraint on the SMC's mass within 3 kpc of $2.29\pm0.46\times10^9\,M_\odot$. Additionally, they found the SMC's inner density profile is consistent with a DM cusp, making the SMC a promising target when searching for DM annihilation signals. Using a separate novel technique, \citet{rathore_response_2025} reported a constraint on the SMC's precollision mass within 2 kpc of (0.8 - 2.4)$\times10^9\,M_\odot$ based on the torque the SMC applies to the LMC's bar during their recent collision. These and future constraints on the SMC's DM mass profile, coupled with extensions to our work in alternative DM models, can pave the way forward for distinguishing between DM particle paradigms.

\subsection{Implications for Isolated Dwarf Galaxy Pairs}\label{subsec:dwarf_pairs}

As our simulations do not include the MW, they can offer general insights into the dynamics of isolated $\sim1:10$ mass-ratio dwarf galaxy pairs. Classical examples of such interacting pairs connected by HI bridges (like the Clouds; \citealt{kerr_magellanic_1957, hindman_low_1963}) include NGC 4490/4485 \citep{clemens_possible_1998}, UGC 9562/9560 \citep{cox_stars_2001}, and NGC 3448/UGC 6016 \citep{noreau_amorphous_1986}. More recently, observational surveys have generated catalogs of interacting dwarf galaxies that are large enough for statistical population studies \citep[e.g.,][]{robotham_galaxy_2012, sales_satellites_2013, stierwalt_tiny_2015, pearson_local_2016, paudel_catalog_2018}. Together with complementary theoretical work on dwarf pairs in cosmological simulations \citep[e.g.,][]{rodriguez-gomez_merger_2015, besla_frequency_2018, chamberlain_fractions_2024, chamberlain_mergers_2024}, it is increasingly clear that interactions between dwarf galaxies are crucial drivers of the star formation histories, morphologies, and baryon cycles of these systems \citep[e.g.,][]{Rakhi2023, Subramanian2024, Hota2024, Geethika2025}. 

We stress that our results are obtained from a single realization of a 1:10 dwarf-dwarf merger, and that details of the host halo's response will be dictated by many factors, including e.g. the exact mass ratio of the interaction, the host's concentration, and the exact orbit of the satellite. Regardless, this work underscores the importance of DM in shaping interactions between dwarf galaxies, and offers guidelines for effects that should be accounted for in studies of these systems.

In particular, a companion can contribute significant amounts of its own DM mass ($\gtrsim50\%$) to the host over a few orbits, as well as induce large-scale distortions in the host halo (wakes, density center displacement). These processes cause large asymmetries in the gravitational potential, invalidating assumptions of equilibrium. Additionally, the morphology of host disks may be influenced by these asymmetries in the halo, in addition to direct tides from the satellite. This motivates the need for a framework for disentangling these effects (H. Rathore et al. in preparation). 

\section{Conclusions}\label{sec:conclusions}

We have used the \texttt{EXP} code to construct BFEs of the DM distribution in an isolated LMC-SMC \textit{N}-body simulation (created using \texttt{GADGET-4}; see Section \ref{subsec:sims}) for the first time. The SMC's orbit about the LMC is based on Model 2 in \Bref{}, where the SMC is bound to the LMC and completes two orbits within 5.77 Gyr. At this time, the simulation is stopped, corresponding to the time at which the LMC-SMC system is expected to first cross the MW's virial radius (``MW infall''). The MW itself is not included in this analysis. Three expansions are created: (1) the LMC Expansion, (2) the SMC Expansion, (3) and the Bound SMC Expansion, where the latter is the expansion for only DM particles that are bound to the SMC. 

Our primary goals are to study perturbations in the LMC's DM halo induced by the SMC and quantify the evolution of the SMC's DM mass profile and halo shape prior to the MW infall of the LMC-SMC binary. Ultimately, our aim is to develop a generalizable framework to understand the DM distribution in hosts with massive satellites ($\sim$1:10 mass ratio; ranging from MW-LMC to LMC-SMC systems), over multiple pericentric approaches of the satellite.

This paper is the first to quantify the evolution of the DM halos in a simulation of the LMC-SMC interaction. We demonstrate that the LMC and SMC induce highly time-dependent, asymmetric perturbations in each other's halos. This implies that static, spherically symmetric halo models cannot accurately capture the evolution of the system's gravitational potential. At MW infall, the distribution of DM in the simulated LMC-SMC system is as follows.

\begin{enumerate}

    \item \textbf{The SMC induces a local dynamical friction wake in the LMC's halo that extends along the SMC's past orbit and persists over multiple pericenters}. At MW infall, the SMC's dynamical friction wake extends over $\sim20$ kpc and has a peak density contrast of $\sim0.4$ relative to the LMC's initial density profile (see Figure \ref{fig:composite}). The SMC's wake is primarily captured by the LMC's quadrupole, which peaks at SMC pericenters but persists throughout the simulation (see Figures \ref{fig:odens_orders} and \ref{fig:modes}). 
   
    \item \textbf{The SMC displaces the LMC's density center at each pericentric approach, exciting a dipole response in the LMC halo.} At the time of MW infall, there are two identifiable overdensities in the LMC's halo associated with the dipole excitations by the SMC (see Figure \ref{fig:composite}). The Inner Overdensity leads the SMC at $\sim 60$ kpc from the LMC's center and has a density contrast of $\sim0.5$ with respect to the initial density profile of the LMC halo. This forms as the SMC approaches its next (third) pericenter passage. The Outer Overdensity (density contrast $\sim 0.4$ at $\sim$ 100 kpc) formed from the density center displacement of the LMC during the SMC's previous (second) pericenter. 
 
    \item \textbf{The LMC's spherically averaged DM density profile at MW infall is similar to its initial profile (within 5\%), although SMC-induced asymmetries are locally important} (Figure \ref{fig:lmc_profiles}). The power in the LMC Expansion at infall is dominated by the monopole. However, the LMC Expansion does have power in odd harmonics. As such, asymmetries in the LMC's halo produced by the SMC (e.g. the wake) cannot be ignored, and will impact the acceleration field of the combined system (see Figure \ref{fig:force-errors}).   

    \item \textbf{The LMC's halo shape at present-day must be impacted by both the MW and SMC.} Simulations of the LMC-MW interaction \citep{garavito-camargo_hunting_2019} demonstrate that the MW induces a significant global quadrupole and dipole in the LMC halo. We find that the SMC-induced dipole can reach peak power comparable to the dipole induced by the MW (Figure \ref{fig:template_power_lmc}). Given that the SMC has likely collided with the LMC since the system's infall to the MW \citep[e.g.,][]{rathore_response_2025}, the perturbations demonstrated here will likely be stronger at present-day. As such, the LMC halo shape at present-day must be influenced by both the MW and SMC, and detailed MW-LMC-SMC models will be required to predict the present-day shape of the LMC halo. 
  
    \item \textbf{The SMC's DM distribution at MW infall is strongly distorted by LMC tides, where SMC debris extends to $\sim100$ kpc from the LMC's center. 90\% of the SMC's bound mass at infall resides within $\approx10$ kpc.} The SMC loses two thirds of its initial DM mass by the time of MW infall -- at this time, its bound DM halo mass is $4.3\times10^{9}\,\rm{M}_\odot$ within 50 kpc, where its DM density profile drops sharply (see Figures \ref{fig:composite}, \ref{fig:smc_mass}, and \ref{fig:smc_profiles}). The SMC's spherically averaged DM density profile differs significantly from the original Hernquist sphere, with only the inner $\sim4$ kpc remaining relatively undisturbed. This radius is consistent with the SMC's tidal radius as determined from the location of the $L_1$ Lagrange point / acceleration minimum between the Clouds (Figure \ref{fig:fields}). However, the SMC is not in equilibrium at large radii even before encountering MW tides or its recent collision with the LMC. The SMC's mass loss history makes it a promising candidate for tidal-stripping-based studies of DM particle physics. The SMC will have a minimal effect on the development of the LMC's wake in the MW due to its low DM mass ($\approx2\%$ of the LMC's) at MW infall. 
 
    \item \textbf{The Bound SMC's DM halo shape at infall is elongated along the SMC's orbit about the LMC.} The strongest $l>0$ harmonic in the SMC expansion is the quadrupole, which peaks at SMC pericenters due to the LMC's tides. The Bound SMC's dipole is also excited following pericenter passages with roughly half the delay seen in the LMC's dipole peaks (see Figures \ref{fig:modes_smc} and \ref{fig:modes}). 

    \item \textbf{The acceleration field of the LMC-SMC system at MW infall differs by up to 20\% from the accelerations that would be expected from symmetric, static halos within $\sim60$ kpc of the LMC's center.}  Mutually-induced distortions to both the LMC and SMC impact the acceleration field of the combined system and orbit of the SMC in a manner that is not well captured by static models (see Figures \ref{fig:fields}, \ref{fig:force-errors}, and \ref{fig:analytic_orbits}). Accurately modeling orbits of LMC-SMC halo tracers such as the LMC's satellites, stellar streams, stellar halo, the Stream, and hypervelocity stars {\it requires} accounting for the evolution of the DM distribution of the LMC and SMC's halos. This result will only be compounded with the inclusion of MW tides and the recent LMC-SMC collision.
    
\end{enumerate}

Our results have a number of interesting parallels with the LMC-MW interaction, and can offer insights into $\sim1:10$ satellite:host DM distributions more generally. 

\begin{enumerate}

\item \textbf{The properties of the SMC dynamical friction wake are consistent with findings of wakes forming in other $\sim$1:10 mass ratio satellite:host systems}. The density contrast of the SMC's wake at infall is roughly twice as strong as the LMC's wake at present-day in the isotropic MW model of \citet{garavito-camargo_hunting_2019}, although it is typical of the LMC's wake strength in cosmological MW-LMC analogs \citep[][]{arora_shaping_2025, darragh-ford_shaping_2025}.

\item \textbf{In our LMC-SMC simulation, the halos of both galaxies develop quadrupoles that peak in strength during pericentric passages of the satellite, consistent with findings from LMC-MW simulations \citep{lilleengen_effect_2023}.} In the host, the quadrupole captures the satellite's wake and is subdominant relative to the host's dipole. In the satellite, the quadrupole is the dominant $l>0$ mode and arises due to elongation of the halo due to the host's tides. 

\item \textbf{Density center displacement in the host occurs at each pericentric approach of a satellite.} Results from this work, cosmological simulations \citep[][E. ]{gomez_fully_2016, arora_shaping_2025, darragh-ford_shaping_2025}, and idealized LMC-MW simulations \citep{garavito-camargo_quantifying_2021, lilleengen_effect_2023} agree that a satellite will induce a dipole in the host halo at each pericenter passage owing to the displacement of the host's COM. We expect this to be a common feature of DM halos in satellite-host ($\sim$1:10 mass ratio) interactions. 
However, the exact orbit of the satellite dictates how the repeatedly excited dipole mode impacts the density structure of the halo.
  
\item \textbf{Kinematic signatures of the wake and density center displacement at MW infall (Figure \ref{fig:rv_shells}):} In velocity space, the SMC's wake induces a $\sim$15\% increase of the tangential velocity dispersion of the LMC's halo particles trailing the SMC. A dipole also appears in the radial velocities of LMC halo particles (relative to the LMC center), indicating a $\sim 15$ km s$^{-1}$ reflex motion of the inner LMC at 30 kpc. These signatures are qualitatively similar to the MW's density center displacement and reflex motion due to the LMC \citep[e.g.,][]{garavito-camargo_hunting_2019}, demonstrating that these signatures are common, but only capture the most recent pericenter passage of a satellite. 

\item \textbf{The DM halos of interacting dwarf galaxies in general are likely to be in disequilibrium.} Satellites in orbit about low mass (dwarf) galaxy hosts will induce halo perturbations (wakes and density center displacements). As such, interacting dwarf galaxy pairs are laboratories to study the dynamical interactions between DM halos, satellites, and the disk of the host. 

\end{enumerate}

We have demonstrated the power of the BFE framework and \texttt{EXP} for characterizing the evolution of a 1:10 mass ratio host:satellite system as the satellite passes through multiple pericenters. In the nearby Universe, satellites of this mass ratio are readily seen \citep[e.g.,][]{zaritsky_satellites_1993, mao_saga_2024}, and in cosmological simulations, \citet{chamberlain_fractions_2024} reported that $\sim 15\%$ of massive galaxies in TNG100 host a 1:4-10 satellite at redshift zero. The features we identify in our LMC-SMC simulation, such as satellite wakes, density center displacement dipole(s), kinematic signatures of satellite-induced perturbations, and disequilibrium of the host halo are also seen in MW-LMC simulations, suggesting they may be widely found in 1:10 interacting pairs and therefore are possibly commonplace in nearby galaxy systems.

The resonant responses in the LMC's halo induced by the SMC over multiple orbits may play an important role in the origin of the irregular shape of the LMC's disk. This motivates an urgent need for a framework to quantify the relative contributions of these resonances compared to the SMC itself in shaping the LMC's disk. More generally, our work shows that understanding the effect of a 1:10 mass ratio satellite on its host's disk must account for the full interaction history of the system. In this way, the halo-disk connection becomes more concrete in the presence of a satellite, providing an exciting path forwards to constrain the nature of DM. 

\begin{acknowledgments}

We thank the anonymous reviewer for an insightful report, which improved the framing of the paper and the presentation of our results. HRF is grateful for productive discussions with Kathryn Johnston, Arpit Arora, Kathryne Daniel, Peter Behroozi, Arjun Dey, and Dennis Zaritsky. 

HRF and GB are supported by NSF CAREER AST-1941096 and NASA ATP award 80NSSC24K1225. HR and GB acknowledge support from NASA FINESST 80NSSC24K1469. FAG acknowledges support from the ANID BASAL project FB210003, from the ANID FONDECYT Regular grants 1251493 and from the HORIZON-MSCA-2021-SE-01 Research and Innovation Programme under the Marie Sklodowska-Curie grant agreement number 101086388. CL acknowledges funding from the European Research Council (ERC) under the European Union’s Horizon 2020 research and innovation programme (grant agreement No. 852839) and funding from the Agence Nationale de la Recherche (ANR project ANR-24-CPJ1-0160-01).

We respectfully acknowledge the University of Arizona is on the land and territories of Indigenous peoples. Today, Arizona is home to 22 federally recognized tribes, with Tucson being home to the O’odham and the Yaqui. Committed to diversity and inclusion, the University strives to build sustainable relationships with sovereign Native Nations and Indigenous communities through education offerings, partnerships, and community service.

The Theoretical Astrophysics Program (TAP) at the University of Arizona provided resources to support this work. This material is based upon High Performance Computing (HPC) resources supported by the University of Arizona TRIF, UITS, and Research, Innovation, and Impact (RII) and maintained by the UArizona Research Technologies department. We thank Ethan Jahn for his assistance with installing and troubleshooting \texttt{EXP} and \texttt{MakeGalaxy} on the Puma cluster, which was made possible through University of Arizona Research Technologies Collaborative Support program. We have made extensive use of NASA's Astrophysics Data System and the arXiv pre-print service in the preparation of this paper. 

\end{acknowledgments}

\begin{contribution}

HRF was responsible for the BFE construction and analysis, and wrote and submitted the manuscript. 
HR developed and ran the \textit{N}-body simulation, contributed Section \ref{subsec:sims}, and edited the manuscript. 
GB supervised HRF and HR, obtained funding, provided guidance on the analysis, and edited the manuscript.
NGC edited the manuscript, and provided guidance on the analysis as well as a snapshot of his simulation described in \citet{garavito-camargo_hunting_2019}.
EP contributed the rigid-potential orbits in Section \ref{subsec:comb_exp} and edited the manuscript.
MSP and MDW provided guidance on the analysis and edited the manuscript. 
FAG and CFPL provided significant comments on an early draft of the manuscript and edited the final version. 


\end{contribution}

%

\software{
\texttt{Astropy} \citep{astropy:2013, astropy:2018, astropy:2022};
\texttt{EXP/pyEXP} \citep{petersen_exp_2025};
\texttt{Gagdet-4} \citep{springel_simulating_2021};
\texttt{H5py} \citep{collette_h5pyh5py_2020};
\texttt{Healpy} \citep{gorski_healpix_2005, zonca_healpy_2019};
\texttt{Jupyter} \citep{kluyver_jupyter_2016, granger_jupyter_2021};
\texttt{Matplotlib} \citep{hunter_matplotlib_2007};
\texttt{Numpy} \citep{harris_array_2020};
\texttt{Scipy} \citep{virtanen_scipy_2020};
          }


\appendix

\section{Expansion Validation}\label{apdx:validation}

\begin{figure*}
    \centering
    \includegraphics[width=\textwidth]{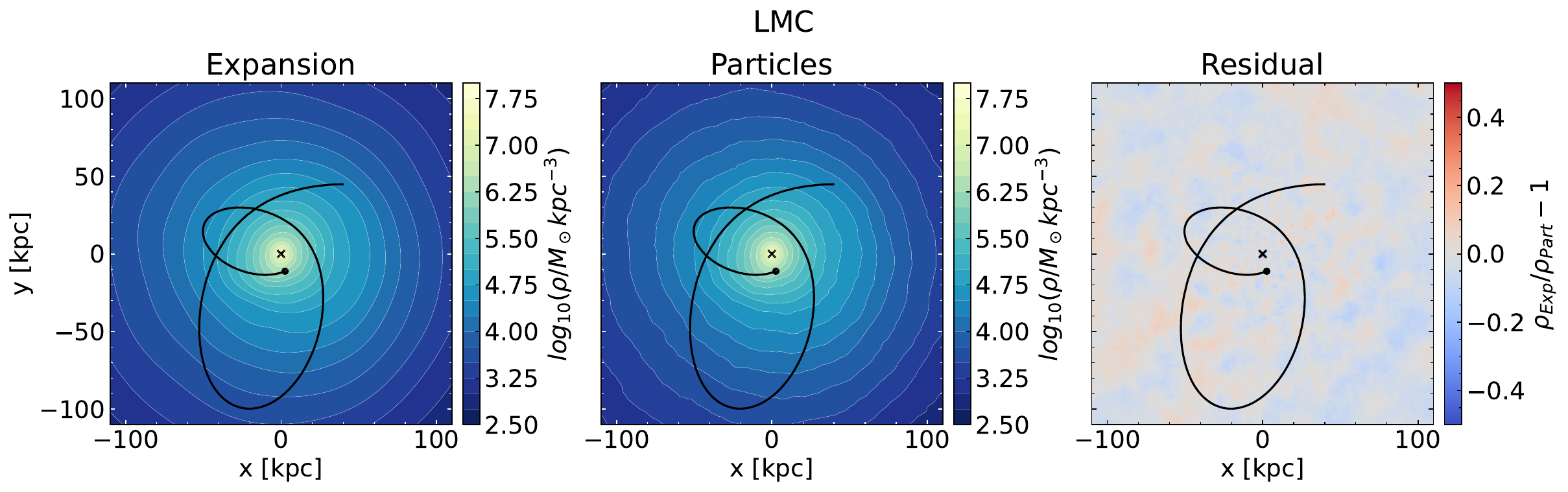}
    \caption{Comparison of the density field reconstructed from the LMC Expansion (\textit{left}) with the density field of the \textit{N}-body particles (\textit{center}) at the MW infall snapshot in the SMC's orbital plane. The right panel shows the relative error $\rho_{Exp}/\rho_{Part}-1$ between the expansion and the simulation particles, demonstrating the LMC Expansion recovers the density field of the LMC to $\lesssim 5\%$.}
    \label{fig:lmc_resid}
\end{figure*}

\begin{figure*}
    \centering
    \includegraphics[width=\textwidth]{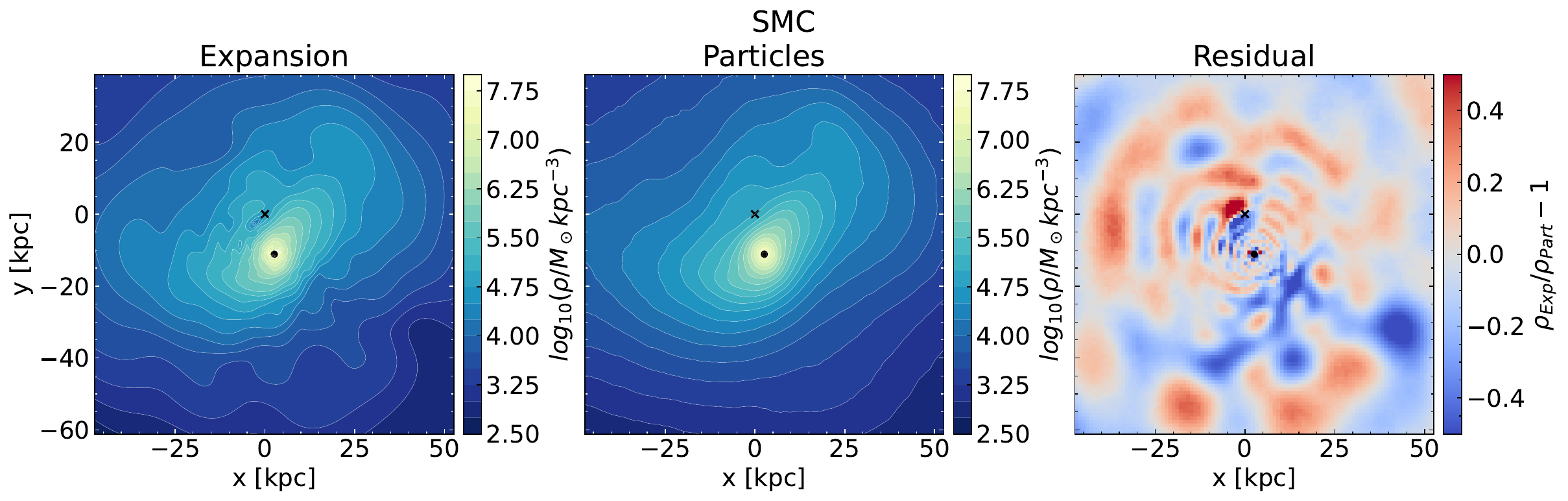}
    \caption{Same as Figure \ref{fig:lmc_resid} for the SMC Expansion. Note each panel is centered on the SMC's center (black point), though the coordinate system remains centered on the LMC (black x). The panel dimensions (100$\times$100 kpc) match the region in which we carry out the MIRSE minimization for this expansion (see Section \ref{subsec:expansions}). While the expansion has large errors ($\gtrsim 40\%$) near the location of the LMC's center and in the outskirts at $y\lesssim20, x>0$, the high-density regions of the SMC ($\rho\gtrsim10^{5.5}\rm{M}_\odot\,kpc^{-3}$) are reproduced to within $\sim10\%$.}
    \label{fig:smc_debris_resid}
\end{figure*}

\begin{figure*}
    \centering
    \includegraphics[width=\textwidth]{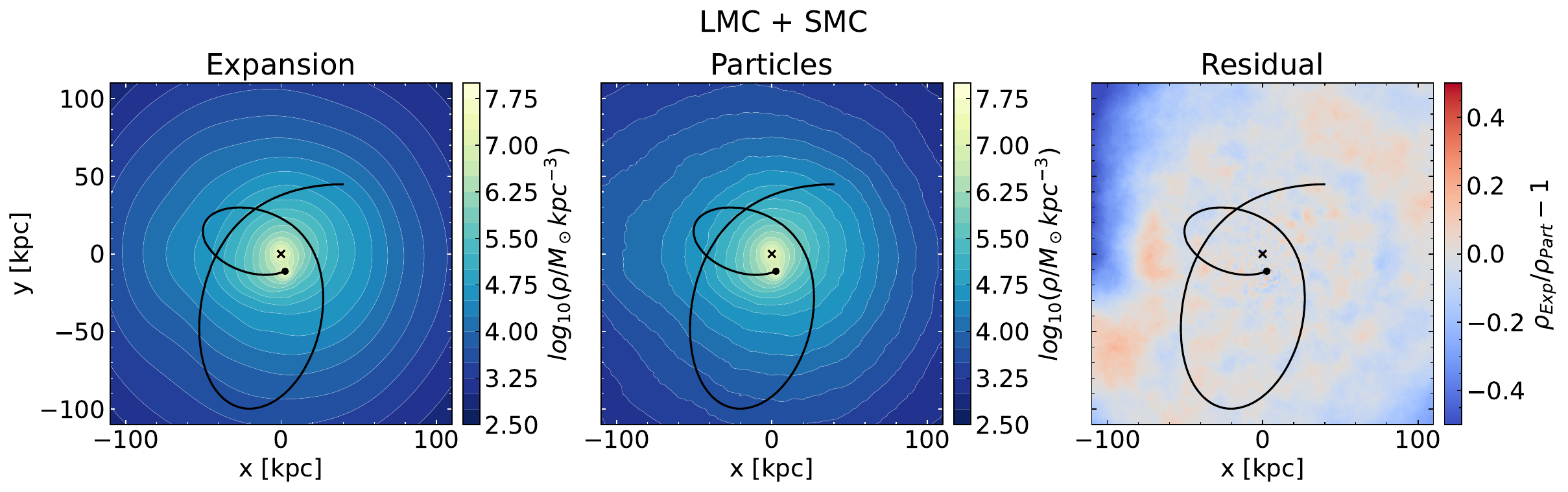}
    \caption{Comparison of the LMC + SMC Expansions (\textit{left)} with the density field of the simulation particles (\textit{center}) at the MW infall snapshot. The right panel shows the relative error between the expansion and the simulation particles, demonstrating the expansions recover the density field of the system to $<10\%$ precision within $\sim 100$ kpc of the LMC's center. This shows that the errors in the SMC expansion seen in Figure \ref{fig:smc_debris_resid} are small compared to the LMC's density and are therefore irrelevant to the combined LMC-SMC DM distribution.}
    \label{fig:sys_resid}
\end{figure*}

\begin{figure*}
    \centering
    \includegraphics[width=\textwidth]{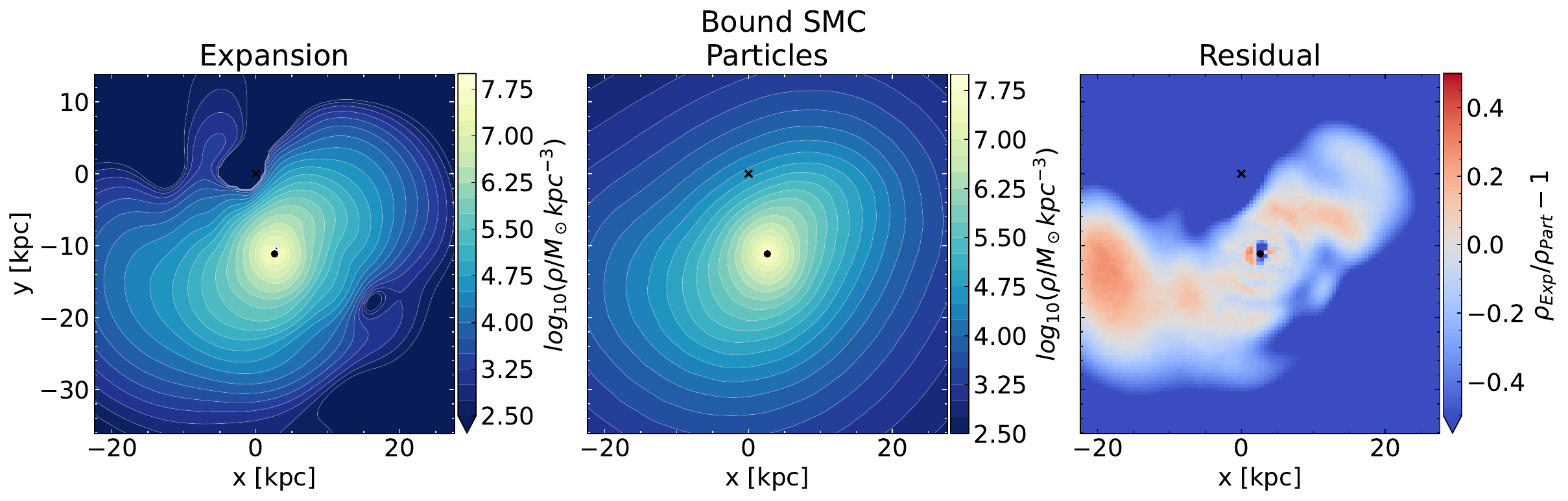}
    \caption{Same as Figure \ref{fig:smc_debris_resid} for the Bound SMC expansion, although note the panel dimensions are 50$\times$50 kpc. The MISE-based truncation for this expansion (see Sections \ref{subsec:truncation} and \ref{subsec:expansions}) reproduces the inner, high-density ($\rho \gtrsim 10^{5.5}\, \rm{M}_\odot kpc^{-3}$) regions of the halo to within $\sim10\%$. Note that the inner 1 kpc of the halo is also not resolved by the basis. In the lower density outskirts, the errors rise to $>50\%$. As we only use this expansion to analyze the shape of the inner Bound SMC, the expansion is acceptable for this purpose.}
    \label{fig:smc_resid}
\end{figure*}

In this section, we compare the density fields reconstructed from our expansions to density fields calculated directly from our \textit{N}-body simulations to quantify the accuracy of our expansions. 

Figure \ref{fig:lmc_resid} compares the LMC's density field as reconstructed from the LMC Expansion to the LMC's density calculated directly from the \textit{N}-body simulation. The LMC expansion is very accurate, recovering the density field from the simulation to sub-5\% accuracy over the entire volume we consider in this work. 

The SMC Expansion is shown in Figure \ref{fig:smc_debris_resid}. The expansion does well at reproducing high-density regions of the SMC, with $<10\%$ errors when the density is above $10^{5.5}\,\rm{M}_\odot\,kpc^{-3}$. However, the expansion fails to accurately describe some of the SMC's debris that has been captured by the LMC near the LMC's center. Additionally, the expansion has $\lesssim40\%$ errors in the outskirts near $y\sim-40$ kpc. For the purposes of tracking the extent of the SMC's debris field, we deem these errors acceptable. 

To ensure that the combination of the LMC and SMC Expansions can accurately recover the combined LMC-SMC density field despite these errors in the SMC Expansion, we compare the sum of the LMC and SMC Expansion density fields to the \textit{N}-body simulation in Figure \ref{fig:sys_resid}. We find that the combined density field is recovered to $<10\%$ within 100 kpc from the LMC's center, showing that the errors in the SMC Expansion are negligible when considering the entire system. 

Figure \ref{fig:smc_resid} compares the Bound SMC Expansion to the simulation. The high-density $\gtrsim 10^{5.5}\,\rm{M}_\odot\,kpc^{-3}$ regions are recovered to $\lesssim10\%$, aside from the inner 1 kpc which the expansion does not resolve. The Bound SMC expansion is therefore accurate for studying the SMC's inner halo. However, the MISE-based truncation of this expansion produces poor results in the outskirts, where errors climb to $>50\%$. 


\bibliography{references}{}
\bibliographystyle{aasjournalv7}



\end{document}